\documentclass[11pt, reqno]{article}

\usepackage{jheppub}
\usepackage{amssymb}
\usepackage{amsmath}
\usepackage[usenames,dvipsnames]{xcolor}
\usepackage{epsfig}
\usepackage{dcolumn}
\usepackage{tikz}
\usetikzlibrary{shapes.geometric, arrows}
\usepackage{upgreek}
\usepackage{setspace}
\usepackage{enumitem}
\usepackage{array,multirow,bigdelim,arydshln}
\usepackage{appendix}
\usepackage{xparse}
\usepackage[utf8]{inputenc}
\hypersetup{
	colorlinks,
	urlcolor=Maroon,
	linkcolor=Maroon,
	citecolor=Maroon
}

\usepackage{float}
\restylefloat{table}

\NewDocumentCommand{\binomial}{omm}
 {%
  \genfrac(){0pt}{}{#2}{#3}%
  \IfValueT{#1}{_{\!#1}}%
 }
\NewDocumentCommand{\eulerian}{omm}
 {%
  \genfrac<>{0pt}{}{#2}{#3}%
  \IfValueT{#1}{_{\!#1}}%
 }

\def \s {\sigma}

\usepackage{underscore}
\newcommand\var{\texttt}

\usepackage{latexsym}
\usepackage{tikz}

\title{Scattering Equations:\\ From Projective Spaces to Tropical Grassmannians}

\author[a]{Freddy Cachazo,}\emailAdd{fcachazo@pitp.ca}
\author[b]{Nick Early,}\emailAdd{earlnick@gmail.com}
\author[a,c,d]{Alfredo Guevara,}\emailAdd{aguevara@pitp.ca}
\author[a,c]{and Sebastian Mizera}\emailAdd{smizera@pitp.ca}

\affiliation[a]{Perimeter Institute for Theoretical Physics, Waterloo, ON N2L 2Y5, Canada}
\affiliation[b]{Massachusetts Institute of Technology, Cambridge, MA, United States}
\affiliation[c]{Department of Physics \& Astronomy, University of Waterloo, Waterloo, ON N2L 3G1, Canada}
\affiliation[d]{CECs Valdivia \& Departamento de F\'isica, Universidad de Concepci\'on, Casilla 160-C,\\ Concepci\'on, Chile}

\abstract{We introduce a natural generalization of the scattering equations, which connect the space of Mandelstam invariants to that of points on ${\mathbb{CP}^1}$, to higher-dimensional projective spaces $\mathbb{CP}^{k-1}$. The standard, $k=2$ Mandelstam invariants, $s_{ab}$, are generalized to completely symmetric tensors $\textsf{s}_{a_1a_2\ldots a_k}$ subject to a `massless' condition $\textsf{s}_{a_1a_2\cdots a_{k-2}\,b\,b}=0$ and to `momentum conservation'. The scattering equations are obtained by constructing a potential function and computing its critical points. We mainly concentrate on the $k=3$ case: study solutions and define the generalization of biadjoint scalar amplitudes. We compute all `biadjoint amplitudes' for $(k,n)=(3,6)$ and find a direct connection to the tropical Grassmannian. This leads to the notion of $k=3$ Feynman diagrams. We also find a concrete realization of the new kinematic spaces, which coincides with the spinor-helicity formalism for $k=2$, and provides analytic solutions analogous to the MHV ones.}

\begin{document}
\maketitle
\addtocontents{toc}{\protect\setcounter{tocdepth}{1}}
\def \tr {\nonumber\\}
\def \nn {\nonumber}
\def \la {|}
\def \ra {|}
\def \dd {\Theta}
\def\hset{\texttt{h}}
\def\gset{\texttt{g}}
\def\sset{\texttt{s}}
\def \be {\begin{equation}}
\def \ee {\end{equation}}
\def \ba {\begin{eqnarray}}
\def \ea {\end{eqnarray}}
\def \k {\kappa}
\def \h {\hbar}
\def \r {\rho}
\def \l {\lambda}
\def \be {\begin{equation}}
\def \en {\end{equation}}
\def \bes {\begin{eqnarray}}
\def \ens {\end{eqnarray}}
\def \red {\color{Maroon}}
\def \pt {{\rm PT}}
\def \s {\textsf{s}}
\def \t {\textsf{t}}
\def \ls {{\rm LS}}
\def \ma {\Upsilon}
\def \SL {{\rm SL}}
\def \GL {{\rm GL}}

\numberwithin{equation}{section}

\section{Introduction: Generalizing the Potential Function}

Scattering equations connect the space of Mandelstam invariants of $n$ massless particles to that of $n$ points on $\mathbb{CP}^1$ \cite{Cachazo:2013gna}. They can be obtained as the conditions for finding the critical points of the ``potential" function
\be\label{ene}
{\cal S} := \sum_{1\leq a<b\leq n} s_{ab}\log\,(a~b).
\ee
Here $s_{ab}$ are known as Mandelstam invariants and $(a~b)$ denotes the ${\rm SL}(2,\mathbb{C})$-invariant combination of the homogeneous variables of points $a$ and $b$. More explicitly,
\be
(a~b):=  \left|
                       \begin{array}{cc}
                         \sigma_{a,1} & \sigma_{b,1} \\
                         \sigma_{a,2} & \sigma_{b,2} \\
                       \end{array}
                     \right|.
\ee
For ${\cal S}$ to be a function of points on $\mathbb{CP}^1$, it has to be invariant under the equivalence relation of projective space $(\sigma_{a,1},\sigma_{a,2})\sim t_a (\sigma_{a,1},\sigma_{a,2})$ for every $a$. This is achieved provided
\be\label{phy}
\sum_{\substack{b=1\\ b\neq a}}^ns_{ab} = 0 \qquad \forall a.
\ee
These conditions have the physical interpretation of momentum conservation. Note that we have not used the diagonal components $s_{aa}$, which in physical applications are taken to be zero for massless particles.

Working in inhomogeneous coordinates $x_a := \sigma_{a,2}/\sigma_{a,1}$ critical points of ${\cal S}$ are found by requiring
\be
\frac{\partial {\cal S}}{\partial x_a} =\sum_{\substack{b=1\\ b\neq a}}^n\frac{s_{ab}}{x_a-x_b} = 0\qquad \forall a,
\ee
which are known as the scattering equations \cite{Cachazo:2013gna}.

The scattering equations are at the heart of the Cachazo--He--Yuan (CHY) formulation \cite{Cachazo:2013hca,Cachazo:2013iea} of scattering amplitudes of a large variety of theories and have made manifest properties such as the Kawai--Lewellen--Tye (KLT) \cite{Kawai:1985xq} and Bern--Carrasco--Johansson (BCJ) relations \cite{Bern:2008qj}. This is one of the motivations for finding natural generalizations. Moreover, generalizations to spaces other than $\mathbb{CP}^1$ can help in better understanding the original equations and in finding new physical applications.

For instance, a large generalization of scattering equations was introduced by one of the authors in \cite{Mizera:2017rqa} and later used to study algebraic properties of multi-loop Feynman integrals \cite{Mastrolia:2018uzb,Frellesvig:2019kgj}. Earlier exploration into constructing a higher ``M\"obius spin'' extension of scattering equations was taken in \cite{Dolan:2014ega}.

In this work we consider another natural generalization: from points on $\mathbb{CP}^{1}$ to points on $\mathbb{CP}^{k-1}$. The standard case has been chosen to correspond to $k=2$ for historical reasons. In physical applications the scattering equations are used to integrate functions on the moduli space known as Parke--Taylor functions \cite{Parke:1986gb}, which are the simplest examples of what are known as leading singularities \cite{ArkaniHamed:2012nw}. In recent work \cite{Cachazo:2018wvl}, we considered generalization of leading singularities from $\mathbb{CP}^{1}$ to higher-dimensional projective spaces in terms of so-called $\Delta$-algebras. Such higher-$k$ leading singularities were first introduced by Franco et al. \cite{Franco:2015rma} and can be used to construct general non-planar on-shell diagrams in ${\cal N}=4$ super Yang--Mills \cite{Bourjaily:2016mnp}.

On $\mathbb{CP}^{k-1}$ the corresponding ${\rm SL}(k,\mathbb{C})$ invariants are determinants of the homogeneous coordinates of $k$ points $( a_1,a_2,\ldots ,a_k )$. It is then natural to introduce the potential function
\be\label{gen}
{\cal S}_k := \sum_{1\leq a_1<a_2\cdots <a_k\leq n} \s_{a_1a_2\cdots a_k}\log\,( a_1,a_2,\ldots ,a_k).
\ee
Once again, requiring ${\cal S}_k$ to be independent of the scaling of each point imposes conditions on the generalized Mandelstam invariants $\s_{a_1a_2\cdots a_k}$. It is easiest to express the condition by completing $\s_{a_1a_2\cdots a_k}$ into a symmetric rank $k$ tensor,
\be\label{cond}
\sum_{\substack{a_2,a_3,\ldots ,a_k=1\\ a_i\neq a_j}}^n\s_{a_1a_2\cdots a_{k}} = 0 \qquad \forall a_1.
\ee
It is tempting to think about the tensor $\s_{a_1a_2\cdots a_{k}}$ as the multiparticle generalization of $s_{ab}$ given by the norm of the sum over the corresponding momentum vectors. As it turns out, while multiparticle invariants satisfy \eqref{cond}, they are not the most general solution and therefore we do not specialize to them as doing so would lead to singular configurations. In order to make this manifest we use a different font. Of course when $k=2$ we can write $\s_{a_1a_2}=s_{a_1a_2}$.

The $\mathbb{CP}^{k-1}$ scattering equations are then given by the conditions for finding critical points of ${\cal S}_k$,
\be\label{scattering-equations}
\frac{\partial {\cal S}_k}{\partial x_a^{(i)}} = 0 \qquad \forall (a,i),
\ee
where $x_a^{(i)}$ represent inhomogeneous coordinates of the puncture $a$ on $\mathbb{CP}^{k-1}$. In this work we initiate the study of these equations. 

We denote the moduli space on which the scattering equations \eqref{scattering-equations} are defined by $X(k,n)$. It can be written as a quotient of the Grassmannian $G(k,n)$ by the $n$-torus action,
\be
X(k,n) := G(k,n)/(\mathbb{C}^\ast)^n.
\ee
Since the diagonal torus action is redundant, the complex dimension of $X(k,n)$ turns out to be $(k{-}1)(n{-}k{-}1)$. Understanding the boundary structure of $X(k,n)$ in general proves to be a difficult mathematical problem \cite{kapranov1993chow,sekiguchi1994versal,sekiguchi1994cross,sekiguchi1997w,sekiguchi1999configurations,sekiguchi2000cross,keel2004chow,keel2006}. Hence we focus mainly on the case $k=3$ with $n \leq 6$.

We start in Section~\ref{sec:scattering-equations} by defining the analogues of the CHY formulae for $k=3$ and computing the associated ``amplitudes'' based on a natural generalization of Parke--Taylor factors. We find that they are rational functions with poles in the kinematic invariants $\s_{ijk}$, as well as more complicated linear combinations of $\s_{ijk}$. We then proceed by identifying singular configurations of points on $\mathbb{CP}^2$ associated to these kinematic poles, for example when multiple points become simultaneously collinear. In Section~\ref{sec:duality} we show that there exists a duality between scattering equations on $X(k,n)$ and $X(n{-}k,n)$, which follows from the corresponding duality on the Grassmannian.

In the case of $k=3$ we find a surprising relation to the so-called \emph{tropical Grassmannian} \cite{speyer2004tropical}, which seems to govern the space of kinematic singularities allowed by the CHY formalism on $X(3,6)$. This allows us to associate a set of ``Feynman diagrams'' to each planar ordering $\alpha$ of $6$ labels, and understand generalizations of biadjoint scalar amplitudes $m^{(3)}_6(\alpha | \beta)$ as a sum over the diagrams compatible with both permutations $\alpha$ and $\beta$ at the same time. This is described in Section~\ref{sec:tropical}.

In Section~\ref{sec:matrix-k-kinematics} we provide an interpretation of the generalized kinematic invariants $\s_{a_1 a_2 \cdots a_k}$ as coming from variables similar to the standard spinor-helicity formalism for $k=2$ in four dimensions. Using this kinematics we are able to prove that for any $k$ and $n$ there exist special classes of analytic solutions to the scattering equations akin to the MHV and $\overline{\text{MHV}}$ sectors in four dimensions.

In \cite{Cachazo:2016ror} it was shown that there exist large kinematic regions in which all the solutions of the standard scattering equations are real and bounded. We generalize this construction to $k=3$ cases by interpreting (the real part of) ${\cal S}_3$ from \eqref{gen} as a potential for interacting particles on $\mathbb{RP}^{2}$. We discuss limitations of this procedure in predicting the number of solutions of the general-$k$ scattering equations due to the fact that different soft kinematic regimes are separated by new singularities appearing at $k=3$.

We conclude with a discussion of results and future directions in Section~\ref{sec:discussion}. We also include two appendices. In Appendix~\ref{app:Euler-characteristic} we prove the number of solutions to the scattering equation by computing the Euler characteristic of $X(3,6)$, while in Appendix~\ref{app:soft-limits} we give a lower bound for the number of solutions for general $n$ using soft limits.

\section{\label{sec:scattering-equations}Scattering Equations on $\mathbb{CP}^{2}$: Jacobians and Amplitudes}

In this section we specialize to $k=3$ in order to carry out explicit computations and build intuition by developing some of the same tools already known for $k=2$.

Let us simplify the notation by denoting inhomogeneous coordinates on $\mathbb{CP}^{2}$ by $(x,y)$. This means that the potential function is
\be
{\cal S}_3 = \sum_{1\leq a<b<c\leq n}\s_{abc}\log \left|
     \begin{array}{ccc}
       1 & 1 & 1 \\
       x_a & x_b & x_c \\
       y_a & y_b & y_c \\
     \end{array}
   \right|.
\ee
The scattering equations are then
\be\label{sck3}
\sum_{\substack{1 \leq b<c \leq n\\ b,c\neq a}}\frac{\s_{abc}(x_b-x_c)}{\la abc\ra}=0 ,\qquad  \sum_{\substack{1 \leq b<c \leq n\\ b,c\neq a}}\frac{\s_{abc}(y_b-y_c)}{\la abc\ra}=0, \qquad \forall\; a.
\ee
Here we have introduced the shorthand notation $\la abc\ra$ for the determinant in ${\cal S}_3$.

At first sight, these are $2n$ equations for $2n$ variables. However, as it is familiar in the $k=2$ case, these equations are covariant under ${\rm SL}(3,\mathbb{C})$ transformations which is the automorphism group of $\mathbb{CP}^2$. This means that $8$ equations are redundant and that we can use the group to fix the positions of $4$ points (each having two coordinates) to generic, i.e., non-collinear, positions.

This makes it clear that $n\geq 4$ in order to have a stable $\mathbb{CP}^2$, i.e., one in which all the automorphism group is fixed. Recall that when $k=2$, this is equivalent to the statement that at least three points are needed in order to have a well-defined set of equations and develop the CHY formalism.

\subsection{Jacobian Matrix}

The next key ingredient in the CHY formalism of scattering amplitudes is the Jacobian matrix associated with the system of scattering equations. Once again, let us review the $k=2$ case before discussing the $k=3$ case.

The Jacobian matrix is a $n\times n$ symmetric matrix with components $\Phi^{(2)}_{ab}:=\partial^2 {\cal S}_2/\partial x_a\partial x_b$. It is well-known that this matrix has rank $n-3$ due to the ${\rm SL}(2,\mathbb{C})$ action, and therefore one has to defined a reduced determinant
\be
{\rm det}'\Phi^{(2)} :=\frac{{\rm det}\Phi^{(2)ijk}_{pqr}}{V_{ijk}V_{pqr}},
\ee
where the Vandermonde determinants are defined as $V_{ijk} := (x_i{-}x_j)(x_j{-}x_k)(x_k{-}x_i)$ and $\Phi^{(2)ijk}_{pqr}$ is a $(n{-}3)\times (n{-}3)$ matrix obtained by deleting rows $i,j,k$ and columns $p,q,r$ of $\Phi$. It is easy to show that the right-hand side is independent of the choice of rows and columns to delete and this is why ${\rm det}'\Phi^{(2)}$ is well-defined.

Moving on to $k=3$, one has that the Jacobian matrix is a $2n\times 2n$ matrix with a block structure. The $n\times n$ blocks have components $\Phi^{(3)}_{ab}:=\partial^2 {\cal S}_3/\partial x_a\partial x_b$, $\Phi^{(3)}_{n+a,n+b}:=\partial^2 {\cal S}_3/\partial y_a\partial y_b$, $\Phi^{(3)}_{a,n+b}:=\partial^2 {\cal S}_3/\partial x_a\partial y_b$, and $\Phi^{(3)}_{n+a,b}:=\partial^2 {\cal S}_3/\partial y_a\partial x_b$.

Once again the matrix is singular and it has rank $2n{-}8$ due to the $\SL(3,\mathbb{C})$ action. The reduced determinant in this case is obtained in a completely analogous manner,
\be\label{red3}
{\rm det}'\Phi^{(3)} :=\frac{{\rm det}\Phi^{(3)ijkl}_{pqrs}}{V_{ijkl}V_{pqrs}},
\ee
where $\Phi^{(3)ijkl}_{pqrs}$ is the matrix obtained from $\Phi^{(3)}$ by deleting rows $\{i,i{+}n,j,j{+}n,k,k{+}n,l,l{+}n\}$ and columns $\{p,p{+}n,q,q{+}n,r,r{+}n,s,s{+}n\}$. The generalized Vandermonde factors are defined as
\be
V_{ijkl} := \la ijk\ra \la jkl\ra \la kli\ra \la lij\ra.
\ee

\subsection{\label{sec:amplitudes}Generalized Biadjoint Amplitudes}

The simplest scattering amplitudes that admit a CHY representation are those of the biadjoint scalar theory \cite{Cachazo:2013iea}. In this $k=2$ construction the amplitudes depend on the choice of two cyclic orderings $\alpha$ and $\beta$ and are computed as
\be\label{biad}
m_n^{(k=2)}(\alpha|\beta) = \frac{1}{{\rm vol}(\SL(2,\mathbb{C}))}\int \prod_{a=1}^n dx_a \prod_{a=1}^{n}\delta\left(\frac{\partial {\cal S}_2}{\partial x_a}\right) {\rm PT}^{(2)}(\alpha){\rm PT}^{(2)}(\beta)
\ee
where Parke--Taylor functions are defined as
\be
{\rm PT}^{(2)}(12\cdots n) := \frac{1}{(x_1-x_2)(x_2-x_3)\cdots (x_n-x_1)}
\ee
and the pre-factor containing ${\rm vol}(\SL(2,\mathbb{C}))$ is there to indicate that the integral as such is not well-defined and the $\SL(2,\mathbb{C})$ redundancy has to be fixed. This is done by fixing three coordinates, say $x_i,x_j,x_k$ and removing three delta functions, say $p,q,r$. There is a Fadeev--Popov factor that is generated and it is given by $V_{ijk}V_{pqr}$.

Explicitly evaluating the formula on the solutions $x^{(I)}_a$ to the scattering equations gives
\be
m_n^{(2)}(\alpha|\beta) = \sum_{I=1}^{(n-3)!}\left.\left(\frac{1}{{\rm det}'\Phi^{(2)}}{\rm PT}^{(2)}(\alpha){\rm PT}^{(2)}(\beta)\right)\right|_{x_a=x_a^{(I)}}.
\ee
The simplest amplitude is for $n=3$. In that case ${\rm det}'\Phi^{(2)} = 1/V_{123}^2$ and $m_3^{(2)}(\alpha|\beta) = 1$.

The first non-trivial amplitude is for $n=4$. A simple computation reveals that, e.g.,
\be
m_4^{(2)}(1234|1234) = \frac{1}{s_{12}}+\frac{1}{s_{23}}.
\ee
Let us also give a result for $n=5$ as this will be useful in the $k=3$ discussion,
\be
m_5^{(2)}(12345|12345) = \frac{1}{s_{12}s_{34}}+\frac{1}{s_{23}s_{45}}+\frac{1}{s_{34}s_{51}}+\frac{1}{s_{45}s_{12}}+\frac{1}{s_{51}s_{23}}.
\ee
Now we are ready to consider $k=3$ amplitudes. In analogy with the biadjoint theory \eqref{biad} we define
\be\label{biad3}
m_n^{(3)}(\alpha|\beta) := \int d\mu_{3,n} \, {\rm PT}^{(3)}(\alpha){\rm PT}^{(3)}(\beta),
\ee
where 
\be
d\mu_{3,n} := \frac{1}{{\rm vol}(\SL(3,\mathbb{C}))} \prod_{a=1}^n dx_a\,dy_a \prod_{a=1}^{n}\delta\left(\frac{\partial {\cal S}_3}{\partial x_a}\right)\delta\left(\frac{\partial {\cal S}_3}{\partial y_a}\right).
\ee
The $k=3$ Parke--Taylor functions are defined as
\be
{\rm PT}^{(3)}(1,2,\ldots , n) := \frac{1}{\la 123\ra\la 234\ra\cdots \la n12\ra}.
\ee
Once again, the factor ${\rm vol}(\SL(3,\mathbb{C}))$ is there to indicate that the corresponding redundancy must be fixed before attempting the integration. After fixing the redundancy one finds
\be
m_n^{(3)}(\alpha|\beta) = \sum_{I=1}^{{\cal N}_n^{(3)}}\left.\left(\frac{1}{{\rm det}'\Phi^{(3)}}{\rm PT}^{(3)}(\alpha){\rm PT}^{(3)}(\beta)\right)\right|_{x_a=x_a^{(I)},y_a=y_a^{(I)}}.
\ee
Here ${\cal N}_n^{(3)}$ is the number of solutions to the scattering equations while the reduced determinant ${\rm det}'\Phi^{(3)}$ is defined in \eqref{red3}. In the cases of interest in this section the number 
of solutions are ${\cal N}_5^{(3)}=2$ and ${\cal N}_6^{(3)} = 26$. We prove this in Appendix~\ref{app:Euler-characteristic}.

Let us illustrate the definition with examples. Clearly, $m_4^{(3)}(\alpha|\beta) = 1$. The first non-trivial amplitudes are for $n=5$.
We start with this case and then move on to $n=6$.

\subsubsection{Case I: $k=3$ and $n=5$}

Explicit computations show that
\be\label{twothree}
m_5^{(3)}(\alpha|\beta)  = \left.m_5^{(2)}(\alpha | \beta)\right|_{s_{ab}\to \s_{cde}}
\ee
where $\{a,b,c,d,e\}=\{1,2,3,4,5\}$.

For example,
\be
m_5^{(3)}(12345|12345) = \frac{1}{\s_{345}\s_{512}}+\frac{1}{\s_{451}\s_{123}}+\frac{1}{\s_{512}\s_{234}}+\frac{1}{\s_{123}\s_{345}}+\frac{1}{\s_{234}\s_{451}}.
\ee
and
\be
m_5^{(3)}(12345|13524) = 0.
\ee

This correspondence between the $k{=}3$, $n{=}5$ case and the $k{=}2$, $n{=}5$ is not an accident. In Section~\ref{sec:duality}, we prove that solutions and amplitudes for $(k,n)$ map to those for $(n{-}k,n)$ as a consequence of the isomorphism between the Grassmannians $G(k,n)$ and $G(n{-}k,n)$.

\subsubsection{Case II: $k=3$ and $n=6$}

Considering $n=6$ and $k=3$ produces new objects but still with similar features to those of $n=6$ and $k=2$. We have computed several explicit examples.

The first class is the set of orderings orthogonal to the identity or canonical order, i.e., those for which the $k=2$ amplitudes vanish. As it turns out the same amplitudes vanish for $k=3$. Explicitly, introducing the notation $\mathbb{I}=123456$ for the canonical ordering, we have:
\be
m_6^{(3)}(\mathbb{I}|125364) = m_6^{(3)}(\mathbb{I}|124635) =m_6^{(3)}(\mathbb{I}|136425) = 0.
\ee
The second class is that of orderings for which only one Feynman diagram contributes to the $k=2$ case:
\be
m_6^{(3)}(\mathbb{I}|126435) =\frac{1}{\s_{612}\s_{345}(\s_{345}+\s_{346}+\s_{356}+\s_{456})(\s_{561}+\s_{562}+\s_{512}+\s_{612})}.
\ee
There are two interesting features that deserve attention. The first is the appearance of both $\s_{612}$ and $\s_{345}$. If they were standard $k=2$ three-particle kinematic invariants momentum conservation would have equated both of them. Here however they are independent and as discussed in the next section, the presence of both poles corresponds to two different geometric configurations; one in which $6,1,2$ are collinear and the other where $3,4,5$ are collinear. The second feature is the structure of the other two poles. Note that they are the straightforward generalization of $(k{+}1)$-particle invariants when $k$ particle invariants are taken as fundamental. Motivated by this we define
\be\label{highK}
\t_{a_1a_2\ldots a_{k}a_{k+1}} := \sum_{i=1}^{k+1} \s_{a_1a_2\ldots \hat{a}_i\ldots a_{k+1}}.
\ee
For example, when $k=2$ one has $t_{abc} = s_{bc}+s_{ac}+s_{ab}$, recall that we use the standard font for both $t$ and $s$ when $k=2$. Using this notation one has
\be
m_6^{(3)}(\mathbb{I}|126435) =\frac{1}{\s_{612}\s_{345}\t_{3456}\t_{5612}}
\ee
and
\be
m_6^{(3)}(\mathbb{I}|125463) =\frac{1}{\s_{123}\s_{456}\t_{6123}\t_{3456}}.
\ee
There is one more topology of $k=2$ amplitudes with a single Feynman diagram. It corresponds to the orderings $(\mathbb{I}|125634)$. This turns out to introduce a new class of poles with no $k=2$ analog. This is why we first discuss other amplitudes which contain two Feynman diagrams in $k=2$ before returning to  $(\mathbb{I}|125634)$,

\be
m_6^{(3)}(\mathbb{I}|125436) =\frac{1}{\s_{345}\s_{126}}\left( \frac{1}{\t_{2345}}+\frac{1}{\t_{3456}}\right)\left( \frac{1}{\t_{5612}}+\frac{1}{\t_{6123}}\right),
\ee

\be
m_6^{(3)}(\mathbb{I}|126453) =-\frac{1}{\t_{3456}\t_{1236}}\left( \frac{1}{\s_{345}}+\frac{1}{\s_{456}}\right)\left( \frac{1}{\s_{612}}+\frac{1}{\s_{123}}\right),
\ee

\be
m_6^{(3)}(\mathbb{I}|145623) =\frac{1}{\t_{1456}\s_{123}\s_{456}}\left( \frac{1}{\t_{6123}}+\frac{1}{\t_{1234}}\right).
\ee

Let us now turn to $m_6^{(3)}(\mathbb{I}|125634)$ to study the new feature of $k=3$ amplitudes. Explicit computation reveals that
\be\label{mist}
m_6^{(3)}(\mathbb{I}|125634) = \frac{\t_{1234}+\t_{3456}+\t_{5612}}{\t_{1234}\t_{3456}\t_{5612}R {\tilde R}},
\ee
where both $R$ and $\tilde{R}$ are polynomials of degree one in the kinematic invariants and are given by
\be\label{R}
R = \s_{123}+\s_{124}+\s_{125}+\s_{126}+\s_{134}+\s_{234}, \quad {\tilde R} = R_{(12)\leftrightarrow (34)}.
\ee
Alternatively, we can denote both combinations by $R_{12,34,56}$ and $R_{34,12,56}$. This notation exhibits the symmetry of the object under cyclic shift of these pairs, i.e.,
\be
R= R_{12,34,56}=R_{34,56,12}=R_{56,12,34}\,.
\ee
These identities require the use of momentum conservation.

The form of $\eqref{mist}$ suggest that the expression should be expanded in terms of three objects; each containing poles of the form $\t \t R\tilde R$. However, it turns out that \eqref{mist} proves to be a single object. In order to motivate this interpretation it is important to note the identity  
\be\label{RRsplit}
R+\tilde{R}=\t_{1234}+\t_{3456}+\t_{5612}.
\ee
This means that \eqref{mist} can also be written as the sum over only two terms, each with a pole structure $\t\t\t R$. This $3$ equals $2$ identity is reminiscent of identities found by computing the ``volume'' of a bipyramid. The volume can be computed either by summing over the top and bottom tetrahedra or by slicing through the middle line and produce three tetrahedra, each containing the top and bottom vertices. This means that none of the two ways of computing the volume are fundamental and that the object of interest is the whole bipyramid. In Section~\ref{sec:tropical} we will see that this intuition is correct, which is why we introduce the notation 
\be\label{bypir}
\dd := \frac{\t_{1234}+\t_{3456}+\t_{5612}}{\t_{1234}\t_{3456}\t_{5612}R {\tilde R}}.
\ee

This object completes the lists of possible singularities appearing in $m_6^{(3)}(\alpha|\beta)$. We can now construct other amplitudes containing two or more $k=2$ Feynman diagrams:
\be\label{nomist}
m_6^{(3)}(\mathbb{I}|125643) = -\frac{1}{\t_{1234}\t_{3456}}\left(\frac{1}{\s_{456} R} +\frac{1}{\s_{123}\tilde{R}}
+\frac{1}{\s_{123}\s_{456}} \right) - \dd\,,
\ee
\begin{align}\label{nomist2}
m_6^{(3)}\!(\mathbb{I}|123465) {=} -\!\frac{1}{\t_{1234}}\!\bigg[
&\frac{1}{R}\!
\left(\! \frac{1}{\s_{234}\s_{456}} {+} \frac{1}{\s_{456}\t_{3456}} {+} \frac{1}{\s_{234}\t_{5612}}\!\right)
\!{+}\frac{1}{\tilde{R}}\!
\left(\! \frac{1}{\s_{123}\s_{156}} {+} \frac{1}{\s_{123}\t_{3456}} {+} \frac{1}{\s_{156}\t_{5612}} \!\right)
\nonumber \\
&+\frac{1}{\s_{456}}\left(\frac{1}{ \s_{123}\t_{3456}}+\frac{1}{\t_{1456}\s_{123}}+\frac{1}{\t_{1456}\s_{234}}\right)\nonumber \\
&+\frac{1}{\s_{156}}\left(\frac{1}{\s_{234}\t_{1256}}+
 \frac{1}{\t_{1456}\s_{234}}+\frac{1}{\t_{1456}\s_{123}}\right) \bigg] - \dd.
\end{align}

There is only one more amplitude missing from the set of all possible `biadjoint' amplitudes. The one missing is $m^{(3)}_6(\mathbb{I}|\mathbb{I})$. As it is familiar from $k=2$ amplitudes, this case is the one with the largest number of terms. We postpone its computation to Section~\ref{sec:tropical} where we present the analog of a Feynman diagram computation, i.e., a purely combinatorial argument which turns out to be based on an interesting connection to tropical Grassmannians.

It can be checked that all amplitudes we have presented are invariant under the exchange $\s_{abc}\to \s_{def}$, where $\{a,b,c,d,e,f\}=\{1,2,3,4,5,6\}$. This exchanges $R$ with $\tilde{R}$ and leaves $\t$'s invariant. This property is the $n=6$ analog of the property \eqref{twothree} and, as we will see in Section~\ref{sec:duality}, follows from self-duality of $G(3,6)$.

Finally, the presence of the different types of singularities in the above amplitudes hints at a richer boundary structure of $X(3,n)$ as compared to $X(2,n)$. Therefore we end this section by studying configurations of points on $\mathbb{CP}^2$ that give rise to these singularities.

\subsection{Geometric Interpretation}

Singularity structure of the amplitudes computed with the new scattering equations is governed by the potential function ${\cal S}_k$. To be more precise, the pole locus of the logarithmic $1$-form $d{\cal S}_k$ specifies the boundaries of $X(k,n)$. The only difficulty in studying these boundaries is that not all of them are accessible in the same chart of $X(k,n)$ as blow-ups might be necessary. Therefore the strategy for checking whether a given configuration of points on $\mathbb{CP}^{k-1}$ is a codimension-$1$ boundary is to first change coordinates such that it is approached as $\varepsilon \to 0$ from a generic configuration, and then compute
\be
\text{Res}_{\varepsilon =0 } \left(d{\cal S}_3 \right),
\ee
which gives the corresponding factorization channel. If the above residue is zero or the change of variables is not valid then the configuration is not a codimension-$1$ boundary of $X(k,n)$. Higher-codimension boundaries are easily obtained by intersecting multiple codimension-$1$ boundaries.

Before discussing the singularity structure of $X(3,6)$ let us consider that of $X(3,5)$, which is more familiar because of the duality to ${\cal M}_{0,5} \cong X(2,5)$ (as discussed in Section~\ref{sec:duality}). Naively, there are two types of singularities that are allowed. The first one is when two points, say $a$ and $b$, collide with each other. It can be parametrized by
\be
x_b = x_a + \varepsilon \hat{x}_b, \qquad y_b = y_a + \varepsilon \hat{y}_b,
\ee
which gives
\be
\la a b c\ra = \varepsilon \Big( (x_a \hat{y}_b - y_a \hat{x}_b) + (\hat{x}_b y_c - \hat{y}_b x_c) \Big), \qquad \la acd \ra = \la bcd \ra + {\cal O}(\varepsilon),
\ee
for $c,d \neq a,b$. Fixing $\hat{x}_b$ and changing the variables from $(x_b, y_b)$ to $(\varepsilon, \hat{y}_b)$ gives
\be
d{\cal S}_3 = \frac{d\varepsilon}{\varepsilon} \sum_{c\neq a,b} \s_{abc} + {\cal O}(\varepsilon^0).
\ee
Using momentum conservation we find $\sum_{c\neq a,b} \s_{abc} = \s_{def}$, where $\{d,e,f\}$ are the three points in the complement of $\{a,b\}$. Therefore $\s_{def}$ is the factorization channel associated to $x_a$ and $x_b$ colliding. More points cannot collide since there is no change of variables allowing for such a scenario.

Another option is that three points, say $d$, $e$ and $f$, become collinear. Clearly the only vanishing angle bracket is $\la def \ra \sim \varepsilon$, which means that
\be
d{\cal S}_3 = \frac{d\varepsilon}{\varepsilon} \s_{def} + {\cal O}(\varepsilon^0),
\ee
and hence this yields the same type of singularity as two complementary points colliding with each other. It is straightforward to check that two codimension-$1$ boundaries intersect only if their channels $\s_{def}$ share exactly one label, e.g., $\s_{123}$ and $\s_{145}$ are compatible, but $\s_{123}$ and $\s_{124}$ are not. This concludes the description of the boundary structure of $X(3,5)$.

Given the above discussion we move on to studying boundaries of $X(3,6)$, where we have the following classes of singularities.

\subsubsection{Two or Three Punctures Colliding}

For $n{=}6$ we can have two or three punctures colliding with each other (four or more is not allowed as it would be inconsistent with $\SL(3,\mathbb{C})$ invariance). In the first case, when $a$ and $b$ collide, we have $\la abc \ra \sim \varepsilon$ for all other $c$ and hence find the channel
\be\label{st-identity}
\sum_{c\neq a,b} \s_{abc} = \t_{defg},
\ee
where $\{d,e,f,g\}$ is the complement of $\{a,b\}$. In the second case, say $a$, $b$ and $c$ colliding \emph{at the same speed} we find
\be
\la abc \ra \sim \varepsilon^2, \qquad \la abd \ra \sim \la acd \ra \sim \la bcd \ra \sim \varepsilon, \qquad \la ade\ra \sim \la bde\ra \sim \la cde\ra \sim  \varepsilon^0
\ee
for all $d,e \neq a,b,c$. Hence the corresponding factorization channel is
\be
2 \s_{abc} + \sum_{d \neq a,b,c} (\s_{abd} + \s_{acd} + \s_{bcd}) = \s_{fgh}.
\ee
Note the factor of $2$ in front of $\s_{abc}$ in the first term due to faster vanishing of $\la abc \ra$. The resulting channel is simply $\s_{fgh}$ for the complementary set $\{f,g,h\}$ to $\{ a,b,c\}$.

\subsubsection{Three or Four Punctures Becoming Collinear}

Next we consider the singularity in which three or four points become simultaneously collinear. In the first case we only have $\la abc\ra \sim \varepsilon$ when $a$,$b$,$c$ become collinear and hence the singularity is simply $\s_{abc}$. Hence it is the same as the complementary three points colliding. In the second case, when $a,b,c,d$ become \emph{simultaneously} collinear, we have
\be
\la abc \ra \sim \la abd\ra \sim \la acd \ra \sim \la bcd \ra \sim \varepsilon,
\ee
while other brackets stay finite. Hence the corresponding singularity is $\s_{abc} + \s_{abd} + \s_{acd} + \s_{bcd} = \t_{abcd}$, which is the same as in the case of the complementary two punctures colliding.

Indeed, by applying $\SL(3,\mathbb{C})$ transformations one can show equivalence of the two sets of singularities.

\subsubsection{Two Punctures Colliding on a Line}

Finally we have a codimension-$1$ singularity in which two points, say $a$, $b$ collide with each other and \emph{at the same rate} become collinear with two other punctures $c$, $d$. In this case we have the vanishing brackets:
\be
\la abe\ra \sim \la acd\ra \sim \la bcd\ra \sim \varepsilon
\ee
for any $e \neq a,b,c$. Hence we obtain the associated channel
\be\label{R-type}
\sum_{e \neq a,b}\s_{abe} + \s_{acd} + \s_{bcd},
\ee
which is the $R$-type singularity: for example when $(a,b,c,d) = (1,2,3,4)$ the sum in \eqref{R-type} equals to $R$ from \eqref{R}. By a change of the $\SL(3,\mathbb{C})$ frame this singularity is the same as $a$ colliding with $b$, while $c$ simultaneously becomes collinear with $a$ and $d$.

Using $\SL(3,\mathbb{C})$ transformations, we can check the invariance of the $R$-type singularity under cyclic shifts of the pairs $(a,b)$, $(c,d)$, $(e,f)$.

\section{\label{sec:duality}Duality Between $X(k,n)$ and $X(n-k,n)$}

The scattering equations inherit a duality from that of Grassmannians $G(k,n)$ and $G(n-k,n)$. In order to show this it is enough to study the potential function.

Let us start with the potential function \eqref{gen} for $n$ points on $\mathbb{CP}^{k-1}$,
\be
{\cal S}_k = \sum_{1\leq a_1<a_2<\cdots <a_k\leq n} \s_{a_1a_2\cdots a_k}\log\,( a_1,a_2,\ldots ,a_k).
\ee
Combining the invariance of this function under $\SL(k,\mathbb{C})$ transformations and rescaling of individual points one has $\GL(k,\mathbb{C})$ as a subgroup. This means that we can go to a gauge fixing or frame in which the first $k\times k$ submatrix for the $k\times n$ matrix defined by the columns of the points is set to the identity. Once this is done, all the maximal minors of the matrix can be interpreted as the minors of a $(n{-}k)\times n$ matrix in which ${\rm GL}(n-k,\mathbb{C})$ has been used to set the maximal minor of the last $n{-}k$ columns to the identity.

This map identifies the $k\times k$ minor $( a_1,a_2,\ldots ,a_k)$ with the $(n{-}k)\times (n{-}k)$ minor $(b_1,b_2,\ldots ,b_{n-k}) $ where the set $\{b_1,b_2,\ldots, b_{n-k}\}$ is the complement of $\{a_1,a_2,\ldots ,a_k\}$ in $\{1,2,\ldots ,n\}$ which can be denoted as $\overline{a_1,a_2,\ldots ,a_k}$.

Applying this to ${\cal S}_k$ one finds
\be
{\cal S}_k = \sum_{1\leq a_1<a_2<\cdots <a_k\leq n} \s_{a_1a_2\cdots a_k}\log\,( \overline{a_1,a_2,\ldots ,a_k}).
\ee

The goal is to show that this function can also be thought of as a general potential function for $n$ points on $\mathbb{CP}^{n-k-1}$. Given that the number of terms, $\binomial{n}{k}$ also happens to be equal to $\binomial{n}{n-k}$, the sum can be transformed into a sum over the complement sets by writing 
\be\label{pot1}
{\cal S}_k = \sum_{1\leq b_1<b_2<\cdots <b_{n-k}\leq n} \s_{\overline{b_1b_2\ldots b_{n-k}}}\log\,( b_1,b_2,\ldots ,b_{n-k}).
\ee 
All that remains to be shown is that the set of kinematics invariants for $\mathbb{CP}^{n-k-1}$ defined by the identification
\be\label{idenF}
\s_{b_1 b_2 \ldots  b_{n-k}}:=\s_{\overline{b_1b_2\ldots b_{n-k}}}
\ee
is generic and satisfies the conditions
\be
\sum_{b_2,\ldots ,b_{n-k}}\s_{b_1b_2\ldots b_{n-k}} = 0, \quad \forall\, b_1\in \{1,2,\ldots ,n\}.
\ee
Once this is done the full $\SL(n{-}k,\mathbb{C})$ invariance can be restored.

Without loss of generality, let us choose $b_1{=}1$. Using the identification \eqref{idenF} one has to prove that 
\be\label{proof}
\sum_{1<a_1<a_2<\cdots < a_k \leq n}\s_{a_1 a_2\ldots a_{k}} = 0.
\ee
Note that the sum does not include the label $1$.  
In order to prove this property, let us start with the conditions $\s_{a_1a_2\ldots a_{k}}$ are known to satisfy, \eqref{cond}, as valid kinematic variables for $n$ points on $\mathbb{CP}^{k-1}$, i.e.,
\be
C_{a_1}:= \sum_{\substack{a_2,a_3,\ldots ,a_k=1\\ a_i\neq a_j}}^n\s_{a_1a_2\cdots a_{k}} = 0 \qquad \forall a_1.
\ee
Consider the linear combination $C_2+C_3+\ldots + C_n$ and collect terms into two groups. The first contains all terms that involve the label $1$ and the second is the rest. This gives
\be
C_2+C_3+\ldots + C_n = \left( \sum_{a_2,a_3,\ldots ,a_k=1; a_i\neq a_j}^n\s_{1 a_2\ldots a_{k}}  \right) + n\left(\sum_{1<a_1<a_2<\cdots< a_k \leq n}\s_{a_1a_2\ldots a_{n}}\right).
\ee 
The first group of terms on the RHS is nothing but $C_1$ while the second is \eqref{proof}. Since all $C_i=0$ this concludes the argument that \eqref{pot1} with the identification \eqref{idenF} defines a valid potential function ${\cal S}_{n-k}$.

Before closing this section note that this duality already shows that the standard scattering equations which are defined on the space of $n$ point in $\mathbb{CP}^1$ are also equations for $n$ points on $\mathbb{CP}^{n-3}$. This is yet another indication of the importance of filling in the gap for other projective spaces in between. 

\section{\label{sec:tropical}Tropical Grassmannians and Higher-$k$ Feynman Diagrams}

In this section we use a surprising connection between tropical Grassmannians, the space of kinematic invariants, and the singularities that can arise in computing amplitudes using the scattering equations. 

The connection starts with the tropical $G(2,n)$ Grassmannian and the standard space of kinematic invariants, i.e., the $k=2$ case. In this section we follow the construction of Speyer and Sturmfels \cite{speyer2004tropical} and use their notation. In particular, they denote the tropical Grassmannian $G(k,n)$ as ${\cal G}_{k,n}'''$. The map connects the vertices of the tropical Grassmannians with all possible kinematics invariant that can be poles of a Feynman diagram in a $\phi^3$ theory. For example, ${\cal G}_{2,4}'''$ is given by three points. These correspond to $s,t,u$. For ${\cal G}_{2,5}'''$ one has $10$ vertices. Each vertex is associated with a given $s_{ij}$. One more case is needed to reach some generality. Consider ${\cal G}_{2,6}'''$ which has $25$ vertices and are in correspondence with $s_{ij}$ and $t_{ijk}$.  In general ${\cal G}_{2,n}'''$ is known to have $2^{n-1}{-}n{-}1$ vertices labeled by all ways of partitioning the set $[n]:=\{1,2,\ldots ,n\}$ into two sets $A,B$ with $|A|>1$ and $|B|>1$.
This is also familiar in the context of matroid theory.

The analogy goes even beyond the structure of vertices. The edges of ${\cal G}_{2,n}'''$ can be mapped to all possible pairs of consistent poles or factorization of an amplitude. In other words, there is an edge between the vertices $\{A,B\}$ and $\{A',B'\}$ if either $A\subset A'$, or $A\subset B'$ or $B\subset A'$ or $B\subset B'$.

Likewise the correspondence continues all the way to the facets for ${\cal G}_{2,n}'''$. As it turns out, the tropical Grassmannian ${\cal G}_{2,n}'''$ has $(2n-5)!!$ facets which is precisely the number of all possible Feynman diagrams in $\phi^3$ theory.\footnote{Connections between tropical geometry and scattering amplitudes in string theory were previously studied by Tourkine in \cite{Tourkine:2013rda}.}

Motivated by this and by the careful study of ${\cal G}_{3,6}'''$ done in \cite{speyer2004tropical} we propose to extend the analogy to $k=3$ kinematics, biadjoint scalar amplitudes, and the corresponding generalization of Feynman diagrams. 

Let us review the ${\cal G}_{3,6}'''$  results of \cite{speyer2004tropical} to exhibit the surprising connection with the objects found up to now and then use it to give a prescription for the computation of the most complicated of the biadjoint scalar amplitudes, i.e., $m^{(3)}_6(\mathbb{I}|\mathbb{I})$.

The main result of \cite{speyer2004tropical} is that ${\cal G}_{3,6}'''$  consists of $65$ vertices, $550$ edges, $1395$ triangles and $1035$ tetrahedra. In order to explain the structure of the vertices one introduces a basis of $20$ unit vectors for $\mathbb{R}^{\binomial{6}{3}}$ 
labeled by three distinct and unordered indices $e_{ijk}$ that take values in $\{1,2,\ldots ,6\}$. The set of these $20$ vectors is denoted as $E$. The space is then modded out by the linear function $\phi: \mathbb{R}^6 \to \mathbb{R}^{\binomial{6}{3}}$, given by
\be
\phi(w_1,\ldots,w_6) = \sum_{\{i,j,k\}\in E}e_{ijk}(w_i+w_j+w_k) = \sum^6_{i=1} w_i\sum_{\substack{j<k\\ j,k\not=i}} e_{ijk}.
\ee
Setting the image of this function to zero accounts for imposing momentum conservation, thus we can identify the set $E$ with $\{\s_{ijk}\}$ in this work. Moreover, for each $4$-subset of labels one defines \cite{speyer2004tropical}
\be
f_{ijkl} = e_{ijk}+e_{jkl}+e_{kli}+e_{lij}.
\ee
The set of such $15$ vectors is denoted by $F$. Clearly, these correspond to $\t_{ijkl}$. Even more surprising is the fact that Speyer and Sturmfels consider objects with six labels
\be
g_{i_1i_2i_3i_4i_5i_6}= f_{i_1i_2i_3i_4}+e_{i_3i_4i_5}+e_{i_3i_4i_6},
\ee
which map to our $R$ and $\tilde R$. The set of these $30$ vectors is $G$. The collection $E\cup F\cup G$ has $65$ vectors and represent the vertices. From now on, we will not distinguish between the vertices and their corresponding kinematic invariants.

The next step is to define edges. There are six classes of edges whose definition can be read off from \cite{speyer2004tropical}. For instance, the first class is ``EE" and is given by pairs of the type $\{\s_{abc},\s_{ade}\}$ or complementary labels as in $\{\s_{abc},\s_{def}\}$. Inspection of the previous amplitudes reveals that only these combinations appear in $m^{(3)}_6(\alpha|\beta)$! This means we are interested in the set of facets as these will correspond to the new $k=3$ Feynman diagrams.

Tetrahedral facets are sets of four vertices $\{v_1,v_2,v_3,v_4\} \subset E$ with each pair $\{v_i,v_j\}$ being an edge. Speyer and Sturmfels find that the $990$ tetrahedra split into six classes and $15$ four-simplices. The six classes are given by all possible relabellings of the following representatives:
\be
    \centering
    \begin{tabular}{|c|c|}
    \hline
       EEEE  & $\{\s_{123},\s_{345},\s_{561},\s_{246}\} $ \\ \hline
        EEFF1 & $\{\s_{123},\s_{456},\t_{1234},\t_{3456}\} $ \\ \hline
        EEFF2 & $\{\s_{123},\s_{345},\t_{3456},\t_{1256}\}$ \\ \hline
        EFFG & $\{\s_{345},\t_{1256},\t_{3456},R_{12,34,56} \}$ \\ \hline
        EEEG & $\{\s_{123},\s_{561},\s_{345},R_{45,23,61}\}$ \\ \hline
        EEFG & $\{\s_{123},\t_{3456},\s_{345},R_{12,34,56}\}$ \\
        \hline
    \end{tabular}\label{tab:classes}\nonumber
\ee
The remaining $15$ four-simplices are permutations of labels of the facet \be
\{\t_{1234},\t_{3456},\t_{5612},R_{12,34,56},R_{12,56,34}\}.\ee 
This is nothing but the $\dd$ diagram defined in \eqref{bypir}. As explained by the authors, tetrahedra contained in these do not correspond to facets. These tetrahedra precisely correspond to the individual terms after splitting $\dd$ using \eqref{RRsplit}. This is the reason why they always appear combined in our previous examples.

Now we are ready to give a prescription for a combinatorial computation of $m^{(3)}_6(\alpha|\beta)$. We start by listing all possible vertices of ${\cal G}_{3,6}'''$ consistent with the planar ordering. This gives the list
\be
L(\mathbb{I})=\{R=R_{12,34,56},R_{23,45,61},\tilde{R}=R_{34,12,56},R_{45,23,61},\s_{123},\ldots,\s_{612},\t_{1234},\ldots,\t_{6123}\}\,,
\ee
of sixteen elements (by $\ldots$ we denote cyclic shifts). Among the $\binomial{12}{4}$ four-element sets of $L(\mathbb{I})$, we simply pick the ones that correspond to facets of ${\cal G}_{3,6}'''$! There are 46 such matches, each of these corresponds to a Feynman diagram of our amplitude. Finally, we append to the list the contribution from the $\dd$ diagram and its cyclic shift $\bar{\dd}$. This gives a list of $48$ elements we denote by $J(\mathbb{I})$. The sum of all the elements of $J(\mathbb{I})$ corresponds to the $m^{(3)}_6(\mathbb{I}|\mathbb{I})$ amplitude, i.e.,
\be
m^{(3)}_6(\mathbb{I}|\mathbb{I}) = \sum_{\Upsilon \in J(\mathbb{I})} \Upsilon\,.
\ee

Most of the terms were shown in Section~\ref{sec:amplitudes}. However, there are new objects: The class EEEG gives contributions of the type $\frac{1}{\s\s\s R}$ which only appear in the $m^{(3)}_6(\mathbb{I}|\mathbb{I})$ amplitude. Very nicely, as in the $k=2$ case, we can compute $m^{(3)}_6(\alpha|\beta)$ simply by considering the common Feynman diagrams of such orderings. In other words, denoting by $J(\alpha)$ and $J(\beta)$ the corresponding relabellings of $J(\mathbb{I})$, we have
\be
m^{(3)}_6(\alpha|\beta) = \sum_{\Upsilon \in J(\mathbb{\alpha})\cap J(\mathbb{\beta})} \!\!\! \Upsilon \,,
\ee
up to an overall sign. Finally, let us note that not all facets appear in the computation of $m^{(3)}_6(\mathbb{I}|\mathbb{I})$. This indicates that more general integrands are needed. For example, one can check that the EEEE facet $\{\s_{123},\s_{345},\s_{561},\s_{246}\}$ never corresponds to a planar ordering. In fact, it is obtained by integrating
\be
\int\frac{d\mu_{3,6}}{(\la 123\ra  \la 345\ra \la 561\ra \la 246\ra)^2 \la 234\ra \la 456\ra \la 612\ra \la 135\ra} = \frac{1}{\s_{123}\s_{234}\s_{345}\s_{246}}.
\ee 
We leave the complete study of these correspondences to future work. However, before ending this section it is worth mentioning some of the results known in the mathematical literature which can help in the exploration of $k=3$ Feynman diagrams and $k=3$ amplitudes. In 2008, Herrmann et al. \cite{herrmann2009draw} revisited the tropical Grassmannian $G(3,6)$ and carefully studied the tropical Grassmannian $G(3,7)$. In their work they computed all rays that define the corresponding spaces and the combinations that make up the facets. Recall that rays (or vertices when considering the intersection with a unit sphere) are in bijection with the kinematic invariants that make ``propagators" while the facets are proposed to correspond to the new $k=3$ Feynman diagrams. All the data collected in \cite{herrmann2009draw} is posted on the webpage:

\url{www.uni-math.gwdg.de/jensen/Research/G3_7/grassmann3_7.html}

Let us explain in more detail how to translate the tropical Grassmannian $G(3,7)$ data on the webpage to physics. The first important object to consider is the table of rays labeled $R$-vector. This is a list of 721 Pl\"ucker vectors. Each entry gives the coefficients of a vector in $\mathbb{R}^{35}$ in the basis $e_{ijk}$ ordered lexicographically. For example, the first entry is
$$(0,0,0,0,0,0,0,0,0,0,0,0,0,0,0,0,0,0,0,0,0,0,0,0,0,0,0,0,0,0,0,0,0,0,1) $$
and corresponds to the ray generated by $e_{567}$. The translation into physics is simply given again by identifying $e_{ijk}$ with $\s_{ijk}$.

As a second example take the element labelled 70 (or 71 in the list)
$$(0,0,0,0,0,0,0,0,0,0,0,0,0,0,1,0,0,0,0,0,0,0,0,0,1,0,0,0,0,0,1,0,0,1,1)$$
which corresponds to $e_{167}+e_{267}+e_{367}+e_{467}+e_{567}$ and translates to $\s_{167}+\s_{267}+\s_{367}+\s_{467}+\s_{567}$. Using momentum conservation this also equals the more familiar invariant $\t_{12345}$.

The table of 721 rays is separated into six classes. We have translated all classes of rays into physical language and found that
\be
    \centering
    \begin{tabular}{|c|c|}
    \hline
       [0,34]  & $\s_{123}$ \\ \hline
       [35,69] & $\t_{1234}$ \\ \hline
       [70,90] & $\t_{12345}$ \\ \hline
        [91,300] & $\t_{12345}+\s_{456}+\s_{457}$ \\ \hline
        [301,615] & $\t_{1234}+\t_{1256}+\s_{127}$ \\ \hline
        [616,720] & $\t_{12345}+t_{34567}+\t_{56712}$ \\
        \hline
    \end{tabular}\label{tab:classes7}\nonumber
\ee
where the first column gives the labels of the first and last elements in the class while the second column is one of the representatives from which the others are obtained by permutations. 

Finally, we can discuss the facets which are presented in the first table of the webpage and which can be read from the column labeled ``Rays". As expected for $k=3$ and $n=7$ each Feynman diagram must consist of six propagators. Here each element in the list contains six rays. Consider for example the first row. The facet is given by rays $(0,7,14,18,25,27)$. From our table above it is easy to see that this facet only contains propagators of the form $s_{ijk}$ and therefore it is in the EEEEEE class in the notation used in this section. More explicitly, this facet then maps to the $k=3$ Feynman diagram
\be
(0,7,14,18,25,27) \to \frac{1}{\s_{567}\s_{347}\s_{246}\s_{235}\s_{145}\s_{136}}.
\ee

\section{\label{sec:matrix-k-kinematics}Matrix Kinematics and MHV Sectors}

The generalized kinematic invariants $\s_{abc}$ have been treated as abstract objects so far. In this section we explore a generalization of the notion of momentum vectors that gives rise to a special class of kinematic invariants $\s_{abc}$ analogous to four-dimensional kinematics when $k=2$. Moreover, on the support of these kinematics one can identify at least four analytic solutions of the scattering equations for all multiplicity, which are analogous to the four-dimensional MHV and $\overline{\text{MHV}}$ solutions \cite{RobertsThesis,Fairlie:1972zz,Fairlie:2008dg}.

Let us start discussing the case of generic $k$ and then specialize to $k=3$. To each particle we associate two $k$-dimensional complex vectors $\lambda^{(a)}_{\alpha}$ and $\tilde\lambda^{(a)}_{\dot \alpha}$ which generalize the well-known spinor-helicity variables. Here the indices $\alpha,{\dot \alpha}\in \{1,2,\ldots ,k\}$ transform under two copies of $\SL(k,\mathbb{C})$. Using these two vectors one can construct a $k\times k$ matrix $\mathbb{K}^{(a)}_{\alpha\dot{\alpha}}=\lambda^{(a)}_\alpha\tilde\lambda^{(a)}_{\dot \alpha}$. The redundancy $\GL(1) \subset \SL(k,\mathbb{C})\times \SL(k,\mathbb{C})$ in this definition is the direct analog of the little group action, thus we can think of each $\lambda^{(a)}_{\alpha}$ and $\tilde\lambda^{(a)}_{\dot \alpha}$ as living in a projective space $\mathbb{CP}^{k-1}$.

The matrix $\mathbb{K}^{(a)}_{\alpha\dot{\alpha}}$ has rank one and therefore its determinant vanishes. Moreover, any linear combination of fewer than $k$ such matrices has rank smaller than $k$ and therefore vanishing determinant. This motivates the following definition
\be\label{spec}
\s_{a_1a_2\cdots a_k} := \det \left(\mathbb{K}^{(a_1)}+\mathbb{K}^{(a_2)}+\cdots +\mathbb{K}^{(a_k)}\right).
\ee
With this definition it is clear that $\s_{a_1a_2\cdots a_k}$ is completely symmetric in its indices. Moreover, if any label repeats the invariant vanishes, i.e., $\s_{a_1a_2\cdots a_{k-2}\,b\,b} = 0$. Furthermore, for $B=\{b_1,\ldots,b_j\}$ with $j>k$, the Cauchy--Binet theorem states that the determinant decomposes as\footnote{A proof of this is obtained by regarding each $\mathbb{K}^{(a)}=\lambda^{(a)}\wedge\tilde{\lambda}^{(a)} $ as a two-form in $k+k$ dimensions. The determinant of any two-form $\omega$ is the coefficient of the top form $\omega^{\wedge k}:=\frac{1}{k!}\omega\wedge\ldots \wedge \omega$. Hence we can identify $s_{a_1\cdots a_k}$ with  $(\sum_i \mathbb{K}^{(a_i)})^{\wedge k} = \mathbb{K}^{(a_1)}\wedge\ldots \wedge \mathbb{K}^{(a_k)}$ and $t_B$ with $(\sum_i \mathbb{K}^{(b_i)})^{\wedge k}$. The result \eqref{CBth} follows after expanding $t_B$. This also motivates the formula \eqref{kinbracket} and the subsequent manipulations.}
\begin{align}\label{CBth}
\t_B:=\det \left(\mathbb{K}^{(b_1)}+\mathbb{K}^{(b_2)}+\cdots +\mathbb{K}^{(b_j)}\right) &= \sum_{\{a_1,\ldots, a_k\}\subset B} \det \left(\mathbb{K}^{(a_1)}+\mathbb{K}^{(a_2)}+\cdots +\mathbb{K}^{(a_k)}\right) \nonumber \\
&= \sum_{\{a_1,\ldots, a_k\}\subset B} \s_{a_1\cdots a_k}.
\end{align}
Hence for $k=3$ the object $\t_{abcd}$ can be identified as the LHS.
The $k=2$ analogy can actually be taken further as one can write
\be\label{kinbracket}
\s_{a_1a_2\cdots a_k} =\langle a_1\ldots a_k\rangle [a_1 \ldots a_k],
\ee
where $\langle a_1\ldots a_k\rangle$ is the determinant of a $k\times k$ matrix of elements $\Lambda^i_\alpha=\lambda^{(a_i)}_{\alpha}$, and analogously $[ a_1\ldots a_k]$ is the determinant of the matrix of elements $\tilde{\Lambda}^{i}_{\dot{\alpha}}=\tilde{\lambda}^{(a_i)}_{\dot\alpha}$. This can be seen by writing the argument of \eqref{spec} as $\Lambda \tilde{\Lambda}^T$. Formula \eqref{kinbracket} also makes explicit the fact that each $\s_{a_1\cdots a_k}$ is linear in the $\mathbb{K}^{(a)}$'s.

So far the invariants built from \eqref{spec} satisfy properties analogous to the $k=2$ case. Let us then impose the generalized momentum conservation constraints \eqref{cond}. First note that adding all the $n$ conditions gives
\begin{align}
0=\sum_{a_1<\cdots < a_k} \s_{a_1\cdots a_k}= \det \left(\mathbb{K}^{(1)}+\cdots+\mathbb{K}^{(n)}\right)
\end{align}
by \eqref{CBth}. Denoting the sum of all momenta by $Q=\sum_{i=1}^n \mathbb{K}^{(i)}$, this condition is the fact that $Q$ is of rank at most $k{-}1$, or, that $Q$ is the sum of at most $k{-}1$ rank-$1$ matrices:
\be\label{qdecomp}
Q=\mathbb{Q}^{(1)}+ \ldots +\mathbb{Q}^{(k-1)}\, , \qquad \mathbb{Q}^{(i)}_{\alpha \dot{\alpha}}=q^{(i)}_{\alpha}\tilde{q}^{(i)}_{\dot{\alpha}}.
\ee
We can write this as $\sum_{a=1}^n \lambda^{(a)}  \tilde{\lambda}^{(a)} = -\sum_{i=1}^{k-1} q^{(i)}\tilde{q}^{(i)}$ and insert it into the conditions \eqref{cond} to obtain
\begin{align}
0&=\sum_{a_2<\cdots < a_k} \langle a_1\cdots a_k \rangle [a_1 \cdots a_k]\nonumber \\
 &=\langle a_1 q^{(1)} \cdots q^{(k-1)}\rangle [ a_1 q^{(1)} \cdots q^{(k-1)}]\,,\quad \forall a_1\,.
\end{align}
To solve these conditions one can assume without loss of generality that 
\be
0= \langle b \, q^{(1)} \cdots q^{(k-1)}\rangle\,,\quad b=1,\ldots, \left\lceil \frac{n}{2}\right\rceil,
\ee
i.e., we have $\left\lceil \frac{n}{2}\right\rceil $ vectors living in a $k{-}1$ plane, where $\left\lceil \frac{n}{2} \right\rceil$ is the ceiling function. But in the cases we are studying ($n\geq 3$ for $k=2$ and $n\geq5$ for $k=3$) we have $\left\lceil \frac{n}{2}\right\rceil >k{-}1$. This means that either the $\left\lceil \frac{n}{2}\right\rceil$ vectors $\lambda^{(b)}$ are degenerate, which would make the kinematic invariants $\s_{b_1\cdots b_k}$ vanish, or the $k{-}1$ vectors $q^{(i)}$ are degenerate, which would decrease the rank of $Q$ to $k{-}2$. As the first case corresponds to singular kinematics we focus on the second setup, which without loss of generality is obtained by writing \eqref{qdecomp} with
\be
\mathbb{Q}^{(k-1)}=0.
\ee
For $k=2$ this sets $Q=0$ and recovers the standard momentum conservation condition. Specializing to $k=3$, generalized momentum conservation now reads
\be\label{k3con}
\mathbb{K}^{(1)}+\mathbb{K}^{(2)}+\cdots +\mathbb{K}^{(n)} = Q = q\, \tilde{q},
\ee
where we call $q = q^{(1)}$ and $\tilde{q} = \tilde{q}^{(1)}$. At the level of the invariants we can use the familiar manipulations of the $k=2$ case, for instance, for $k=3$, $n=6$ we can write
\be
\t_{defg}= \s_{abQ} = \sum_{c=1}^{6} \s_{abc},
\ee
which is \eqref{st-identity}.

\subsection{Analytic Solutions to the Scattering Equations}

In the $k=2$ case it is well known that solutions of the scattering equations split into $n{-}3$ sectors \cite{Spradlin:2009qr,Cachazo:2013iaa}. The $d$-th sector can be associated with two maps $\rho,\tilde{\rho}:\mathbb{CP}^1\to\mathbb{CP}^1$ of degrees $d$ and $n{-}2{-}d$ respectively. For $d=1$ and $d=n{-}3$ the maps are linear and the corresponding solutions are said to be in the MHV and $\overline{\text{MHV}}$ sectors.

Here we provide evidence of the existence of such sector decomposition for higher $k$. We do this by constructing four analytic solutions which are present at any multiplicity for $k=3$. In particular this proves the existence of solutions to our scattering equations. The first two solutions lie in the direct analog of MHV and $\overline{\text{MHV}}$ sectors and are easy to construct for any $k$. The other two are particular for $k=3$ as all the points in $\mathbb{CP}^2$ are found to lie on a conic. As they can also be identified as MHV-like solutions under a Veronese action, we will refer to them as $\text{MHV}_q$ solutions. For $X(3,5)$ the $\text{MHV}_q$ ($\overline{\text{MHV}_q}$) sector coincides with the MHV ($\overline{\text{MHV}}$) one. In particular this implies that there are only two solutions to the scattering equations at five points instead of four.

Let us first discuss the MHV solutions, the $\overline{\text{MHV}}$ case being obtained by exchanging $\lambda_{\alpha} \leftrightarrow \tilde\lambda_{\dot \alpha}$. A degree-one map $\rho: \mathbb{CP}^{k-1}\to\mathbb{CP}^{k-1}$ can be written as $\rho(\sigma) = G  \sigma$, where 
$G$ can be set to the identity by means of a ${\rm GL}(k,\mathbb{C})$ transformation in $X(k,n)$. Setting $k=3$, we adopt inhomogeneous coordinates by putting $\sigma=t(1,x,y)$. This means we can obtain our MHV solution simply as $\lambda^{(a)} = \sigma_a$, i.e.,
\be\label{MHV}
x_a=\frac{\lambda^{(a)}_2}{\lambda^{(a)}_1}\,,\qquad y_a=\frac{\lambda^{(a)}_3}{\lambda^{(a)}_1},
\ee
where we have also solved the scale as $t_a = \lambda^{(a)}_1$. Inserting $\langle abc\rangle = t_a t_b t_c (abc)$ into the scattering equations gives
\be
\sum_{b,c} \frac{\langle abc \rangle [abc]}{(abc)} x_{bc} =  t_a \sum_{b,c} t_b t_c x_{bc}[abc] =  t_a \sum_{b,c} \langle rbc \rangle [abc] \,,\quad r=(0,0,1)\,,
\ee
and the analogous one for the $y$ coordinates. Momentum conservation \eqref{k3con} turns the last expression into $\langle rqq\rangle [aqq] =0$. This proves \eqref{MHV} is a solution. Note finally that the scale $t_a$ drops in the cross-ratio combinations, which are also ${\rm GL}(k,\mathbb{C})$ invariant,
\be\label{crmhv}
\frac{\langle abf\rangle \langle cdf\rangle }{\langle adf\rangle \langle cbf\rangle} = \frac{( abf)( cdf) }{( adf) ( cbf)}.
\ee

This feature will have a nice analog for the ${\rm MHV}_q$ solutions, which we now introduce. For $k=3$ we define such solutions as the ones lying on a conic in $\mathbb{CP}^2$. As there is always a conic passing through five points, this explains why the ${\rm MHV}_q$ sector is contained in the MHV ones for $X(3,5)$. For $n>5$ one needs to set additional conditions so that the $n$ points lie on the conic defined by any five of them. Such conditions, i.e., the conic equations, can be stated as 
\be\label{conic}
\frac{( abf)( cdf) }{( adf) ( cbf)}=\frac{( abg)( cdg) }{( adg) ( cbg)}\,,\qquad f,g\neq a,b,c,d
\ee
(this is an equation for $\sigma_g$ if $\{\sigma_a, \sigma_b,\sigma_c,\sigma_d,\sigma_f\}$ are considered fixed). It can be shown that these conditions are enough to arrange, by means of a ${\rm GL}(k,\mathbb{C})\times (\mathbb{C}^\ast)^n$ transformation, any element of $X(3,n)$ into the Veronese form \cite{ArkaniHamed:2009dg}. Such form is defined as $y_a= x_a^2$ in inhomogeneous coordinates. Under this parametrization $(abc)=x_{ab}x_{bc}x_{ca}$ and the scattering equations become

\be\label{vereq}
0=\sum_{b,c\neq a} \frac{\langle abc \rangle [abc]}{x_{ab}x_{ac} }\,,\qquad  0=\sum_{b\neq a} \frac{\langle abq \rangle [abq]}{x_{ab} } \qquad \forall a\,.
\ee

As the objects $\langle abq\rangle$ and $[abq]$ can be identified with the standard $k=2$ brackets $\langle ab\rangle$ and $[ab]$ (for instance by choosing a frame where $q=(0,0,1)$), we recognize in the second set of conditions the standard scattering equations over $\mathbb{CP}^1$. As before, we know it admits two MHV-type solutions, which can be stated in a covariant form as
\be\label{MHVq}
x_a=\left\{\begin{matrix} \dfrac{\langle a X q\rangle}{\langle a Y q\rangle } \qquad\text{for }\quad \text{MHV}_q, \\
\\ \dfrac{[ a X q]}{[ a Y q] } \qquad{\text{for  }\quad \overline{\text{MHV}}_q}, \end{matrix}\right.
\ee
where $X,Y$ are two reference vectors parametrizing the ${\rm SL}(2,\mathbb{C})$ redundancy of \eqref{vereq}. As it turns out only these two solutions of the scattering equations are also solutions to the first set of conditions in \eqref{vereq}. To see that such condition holds, let us take the ${\rm MHV}_q$ solution and use the Schouten identity to  write $x_{ab}=\frac{\langle abq\rangle \langle XYq\rangle}{\langle aYq \rangle \langle bYq\rangle}$ so that
\begin{align}
\sum_{b,c\neq a} \frac{\langle abc \rangle [abc]}{x_{ab}x_{ac} } &= \frac{\langle aYq\rangle^2}{\langle XY q\rangle ^2 }\sum_{b,c\neq a} \frac{\langle bYq\rangle \langle
cYq\rangle}{\langle abq\rangle \langle acq\rangle } \langle abc \rangle [abc] \nonumber \\
&=  \frac{\langle aYq\rangle}{\langle XY q\rangle ^2 }\sum_{b,c\neq a} \frac{\langle bYq\rangle \langle
cYq\rangle}{\langle abq\rangle \langle acq\rangle }\big(\langle aYc \rangle  \langle abq \rangle - \langle aYb \rangle  \langle acq \rangle\big )[abc] \nonumber \\
&= 2 \frac{\langle aYq\rangle}{\langle XY q\rangle ^2 }\sum_{b,c\neq a} \frac{\langle bYq\rangle \langle
cYq\rangle}{ \langle acq\rangle } \langle aYc \rangle  [abc] = 0,
\end{align}
where the sum over $b$ again vanishes due to momentum conservation. Thus we have found two new solutions which lie on a conic given by \eqref{conic}. In fact, the cross-ratio in \eqref{crmhv} now becomes $\frac{x_{ab}x_{cd}}{x_{ad}x_{cb}}$ so that we can write
\be
 \frac{\langle abq\rangle \langle cdq\rangle }{\langle adq\rangle \langle cbq\rangle} = \frac{( abf)( cdf) }{( adf) ( cbf)}  \qquad \forall f\,,
\ee
which is a ${\rm GL}(k,\mathbb{C})$-invariant statement.

\section{Positive Kinematics}\label{sec:poskin}

Motivated by the work of Kalousios \cite{Kalousios:2013eca}, Zhang and two of the authors found a subregion of the $n(n{-}3)/2$-dimensional space of kinematic invariants $s_{ab}$ where all $(n{-}3)!$ solutions to the standard scattering equations, i.e., $k=2$, are real \cite{Cachazo:2016ror}. Moreover, the equations had the interpretation of the equilibrium points of a potential describing interacting particles on the interval $[0,1]$. This is easy to see by singling out three particles, $A,B,C$ and using $\SL(2,\mathbb{C})$ to set $x_A=0$, $x_B=1$ and $x_C=\infty$. The potential function is then\footnote{Notice that taking the real value $\text{Re}({\cal S}_k)$ instead of ${\cal S}_k$ does not affect the positions of the critical points if all $s_{a_1 a_2 \cdots a_k}$ are real.}
\be
\text{Re}({\cal S}_2) = \sum_{a=1}^{n-3}\Big(s_{Aa}\log |x_a| +s_{Ba}\log |1{-}x_a|\Big) + \sum_{1\leq a<b\leq n-3} \!\!\!\! s_{ab}\log|x_a{-}x_b|.
\ee
Letting all $s_{Aa}$, $s_{Ba}$, $s_{ab}$ be positive gives rise to a system of $n{-}3$ particles on an interval, where all particles repel each other and are also repelled from the boundaries of the interval at $x{=}0$ and $x{=}1$. This region is called ${\cal K}_{n}^{+}$. Note that the choice $s_{Aa}, s_{Ba}, s_{ab}>0$ is possible because they form a basis of the $n(n{-}3)/2$-dimensional kinematic space.\footnote{For a mathematical perspective, see, e.g., \cite{aomoto1975}.}

In this section we generalize the notion of positive kinematics to $k{=}3$ and discuss some of the new features that appear. In the same way as for $k{=}2$, we find for $n{<}7$ that all solutions are real and give rise to a very elegant and pictorial derivation of the number of solutions. 

We start by selecting four particles $A,B,C,D$ to be fixed by the action of $\SL(3,\mathbb{C})$. This time two particles, say $C,D$, can be sent to infinity by setting their homogeneous coordinates to $(0,1,0)$ and $(0,0,1)$ respectively. The other two are chosen to be, in inhomogeneous coordinates, at the origin and at $(1,1)$ on the plane $(x,y) \in \mathbb{R}^2$, i.e., $A$ has homogeneous coordinates $(1,0,0)$ while $B$ has $(1,1,1)$. 

Clearly the $n=4$ case is trivial as all four particles are gauge fixed. Since interactions in the potential function are controlled by the determinants $\la abc\ra$ a given particle is not directly sensitive to the location of any other particle but only to the lines defined by any pair of particles. In order to find the analog of the positive region in this case let us again consider the potential function 
\be\label{yar}
\text{Re}({\cal S}_3) = \!\!\!\!\!\! \sum_{\substack{a=1\\ I<J\in\{A,B,C,D\}}}^{n-4} \!\!\!\!\!\!\!\!\!\! s_{IJa}\log |IJa| + \!\!\!\!\!\sum_{\substack{1\leq a<b\leq n-4\\ I\in\{A,B,C,D\}}} \!\!\!\!\!\!\! s_{Iab}\log|Iab| + \!\!\!\!\!\!\!\sum_{1\leq a<b<c\leq n-4} \!\!\!\!\!\!\!\! s_{abc}\log |abc|.
\ee
Our first approach to the $k=3$ positive region ${\cal K}_{3,n}^+$ is then to ask all invariants that explicitly appear in \eqref{yar} to be positive. This is possible because once again they form a basis for the space of $k=3$ kinematic invariants. More explicitly, the only constrains on kinematic invariants are determined by the $n$ conditions which generalize $k=2$ momentum conservation. These are linear equations for the $n$ variables $s_{ABC}$, $s_{BCD}$, $s_{ACD}$, $s_{ABD}$ and $s_{CDa}$ for $1\leq a\leq n{-}4$. 

\begin{figure}[!h]
\centering
\includegraphics[scale=.4]{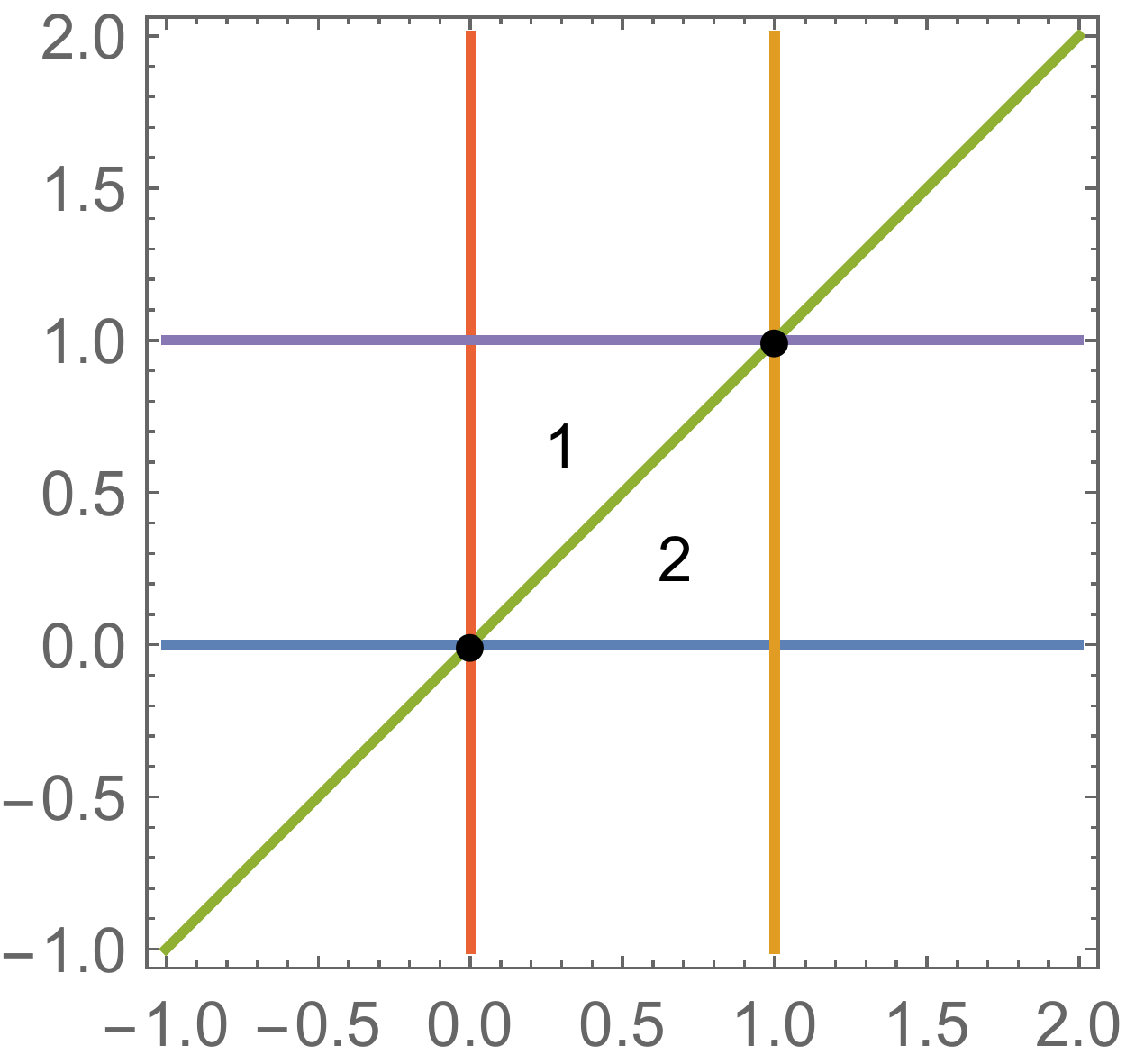}\qquad\qquad
\includegraphics[scale=.4]{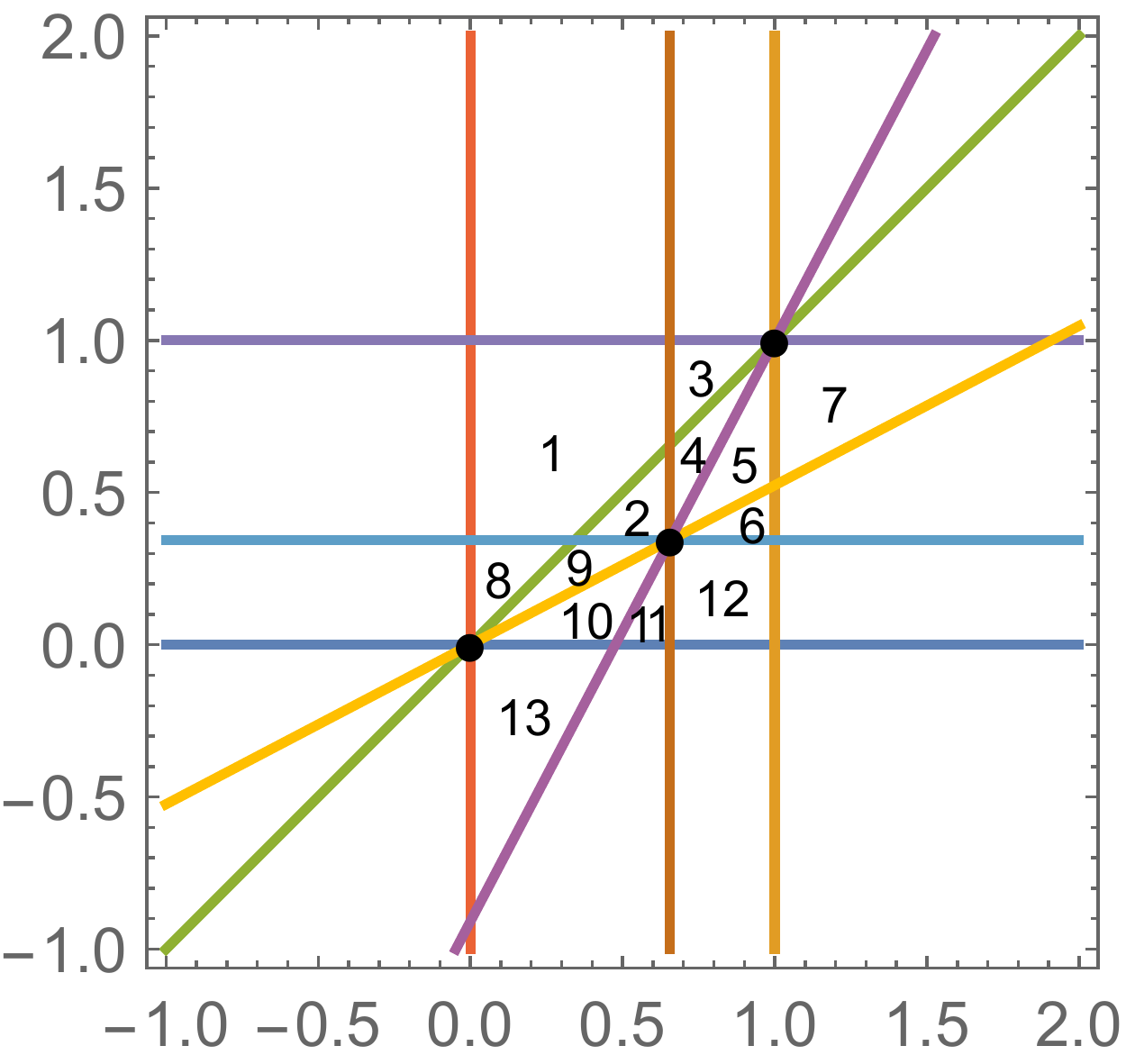}
\caption{\label{fig:chambers}Left: Five lines seen by the free particles when $n=5$. Points $A$ and $B$ are shown as black dots while $C$ and $D$ are at infinity. Out of all $12$ chambers, only two are bounded and are labeled. Right: Once the point $1$ is fixed to a location in the bounded chamber $2$, it generates four new lines which then increase the total number of chambers to $31$. Only $13$ chambers are bounded and are explicitly labeled.}
\end{figure}

Now we are ready to study the dynamics generated by the potential ${\cal S}_3$. Consider $n=5$. Since only one point is free to move we use $x,y$ for its coordinates. Here the potential is 
\be
\text{Re}({\cal S}_3) = s_{AB1}\log |x{-}y| +s_{AC1}\log |y| +s_{AD1}\log |x| +s_{BC1}\log |1{-}y| + s_{BD1}\log |1{-}x|.
\ee
Since all coefficients are positive it is possible to understand the dynamics as that of a particle in $\mathbb{R}^2$ which is repelled from five lines. The lines are, in the order in which they appear in the potential, $x=y$, $y=0$, $x=0$, $y=1$, and $x=1$. This is shown in Figure~\ref{fig:chambers} (left). Since critical points of the potential correspond to equilibrium points, it is clear that they can only lie in the bounded chambers of the space, i.e., those with finite area. Note that the five lines divide the plane into $12$ chambers. Ten of them are unbounded and two are bounded. Therefore there are only two possible places for particle $1$ to be located, and since we know from Section~\ref{sec:amplitudes} that there are exactly two solutions to the scattering equations, they have to lie in these two chambers.

Much more interesting is the $n=6$ case. Here we use the soft limit approach in order to more clearly understand the dynamics. We assume that $|s_{ij2}| \ll |s_{klm}|$. This means that one can effectively solve the scattering equations in steps. First we find solutions for particle 1. As before, we find two of them in the two bounded chambers. Once a solution is picked, we study the dynamics of the second particle, assumed to be soft. Let us choose the lower chamber (chamber $2$ in Figure~\ref{fig:chambers}). Particle $2$ now interacts with the same five lines given above and also with four new lines. The new lines all pass through the point $1$. One is parallel to the $x$ axis, another to the $y$ axis, while the remaining two pass one through $A$ at $(0,0)$ and the other through $B$ at $(1,1)$. These lines are depicted in Figure~\ref{fig:chambers} (right).

Clearly, particle $2$ can only find equilibrium points on bounded chambers. It is easy to count $31$ chambers in total with $13$ bounded and $18$ unbounded. On the figure we have explicitly labeled all $13$ bounded chambers. This can be repeated again for the second equilibrium point of particle $1$ thus obtaining another set of $13$ critical points, summing to $26$. This saturates the total number of solutions (see Appendix~\ref{app:Euler-characteristic}) and hence the above argument describes all of them.

These pictures also reveal that our first attempt at defining a positive region cannot be a completely connected one. The reason is that according to our definition, one can smoothly change the kinematics invariants from the soft region $|s_{ij2}| \ll |s_{klm}|$ to a new one where the roles of $1$ and $2$ are reversed, i.e., $|s_{ij1}| \ll |s_{klm}|$. This leads to a problem as the solutions where particle $2$ is placed on region $7$ or region $13$ in the figure disappear as soon as particle $1$ becomes soft and $2$ hard. 

This means that there must be new singularities that separate the two soft regions. These are not of the form $s_{ijk}=0$ or $t_{ijkl}=0$ since they all have definite sign in the positive region. This puzzle is resolved by the novel $k=3$ singularities found in the biadjoint amplitudes in Section~\ref{sec:amplitudes} and denoted by $R$ and $\tilde R$. 

It is possible to see that some of the $R$'s and $\tilde R$'s do not have definite sign and hence can become zero as we move from one soft region to the next. 

\begin{figure}[!h]
\centering
\includegraphics[scale=.4]{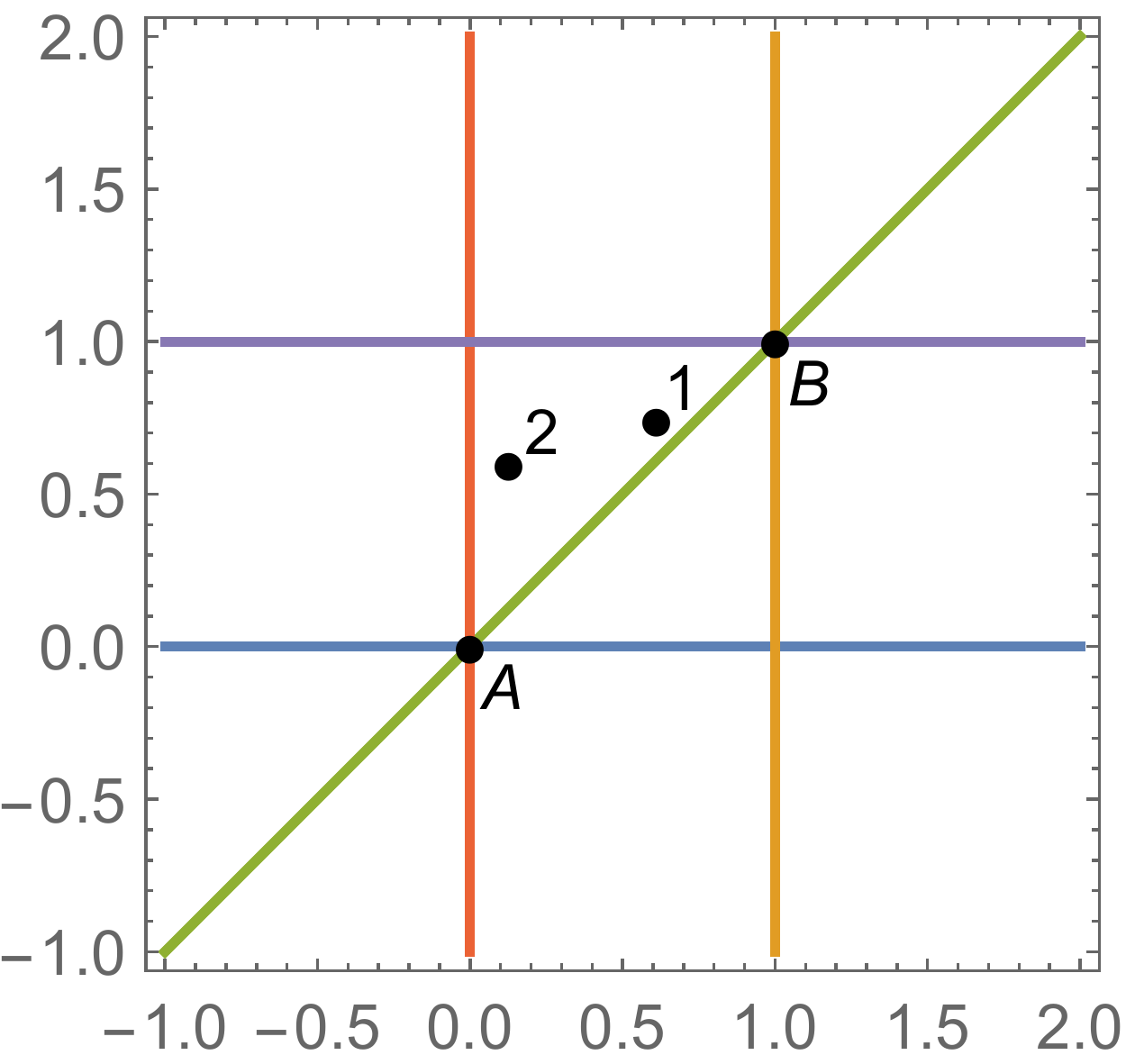}
\includegraphics[scale=.4]{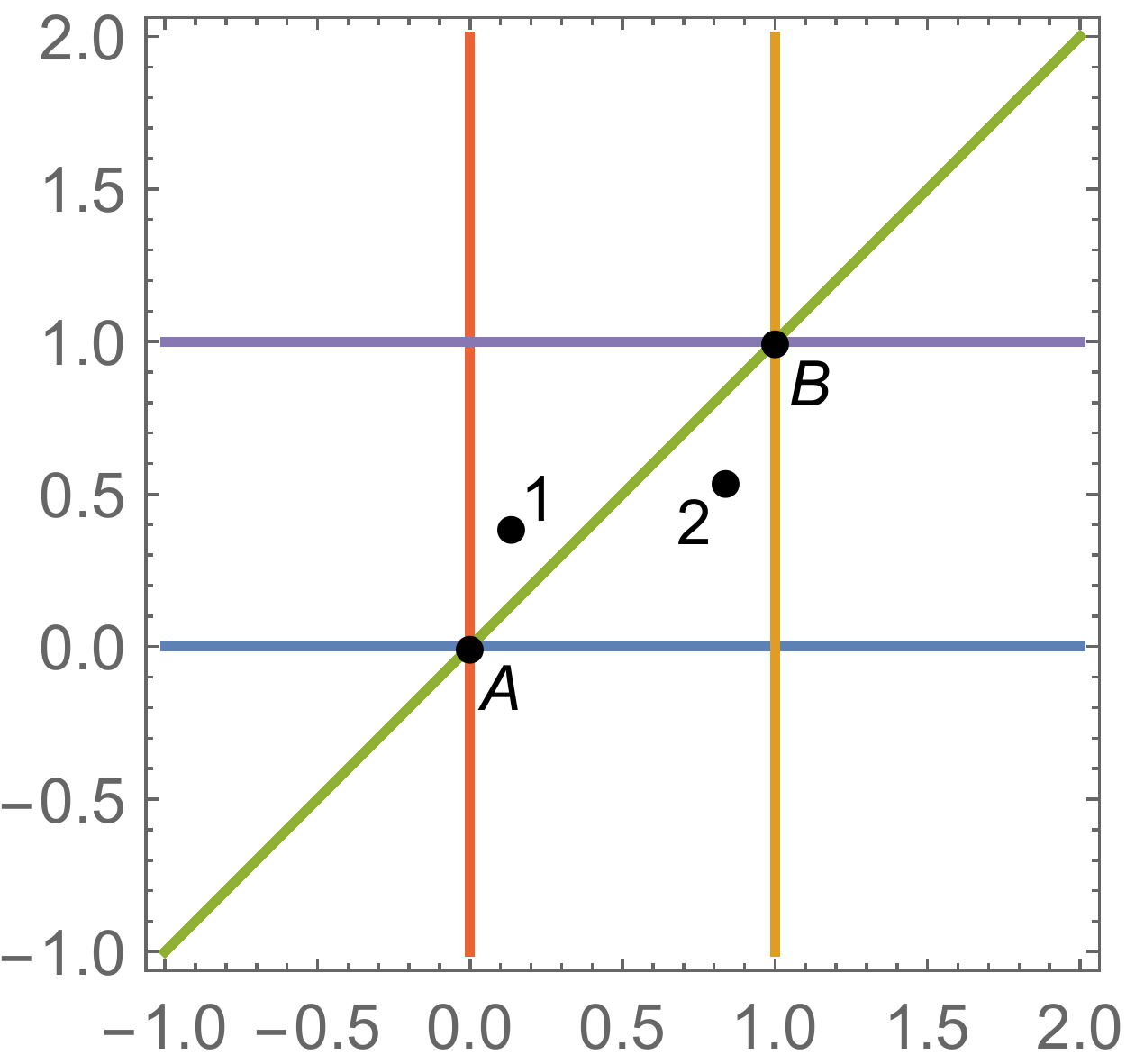}
\includegraphics[scale=.4]{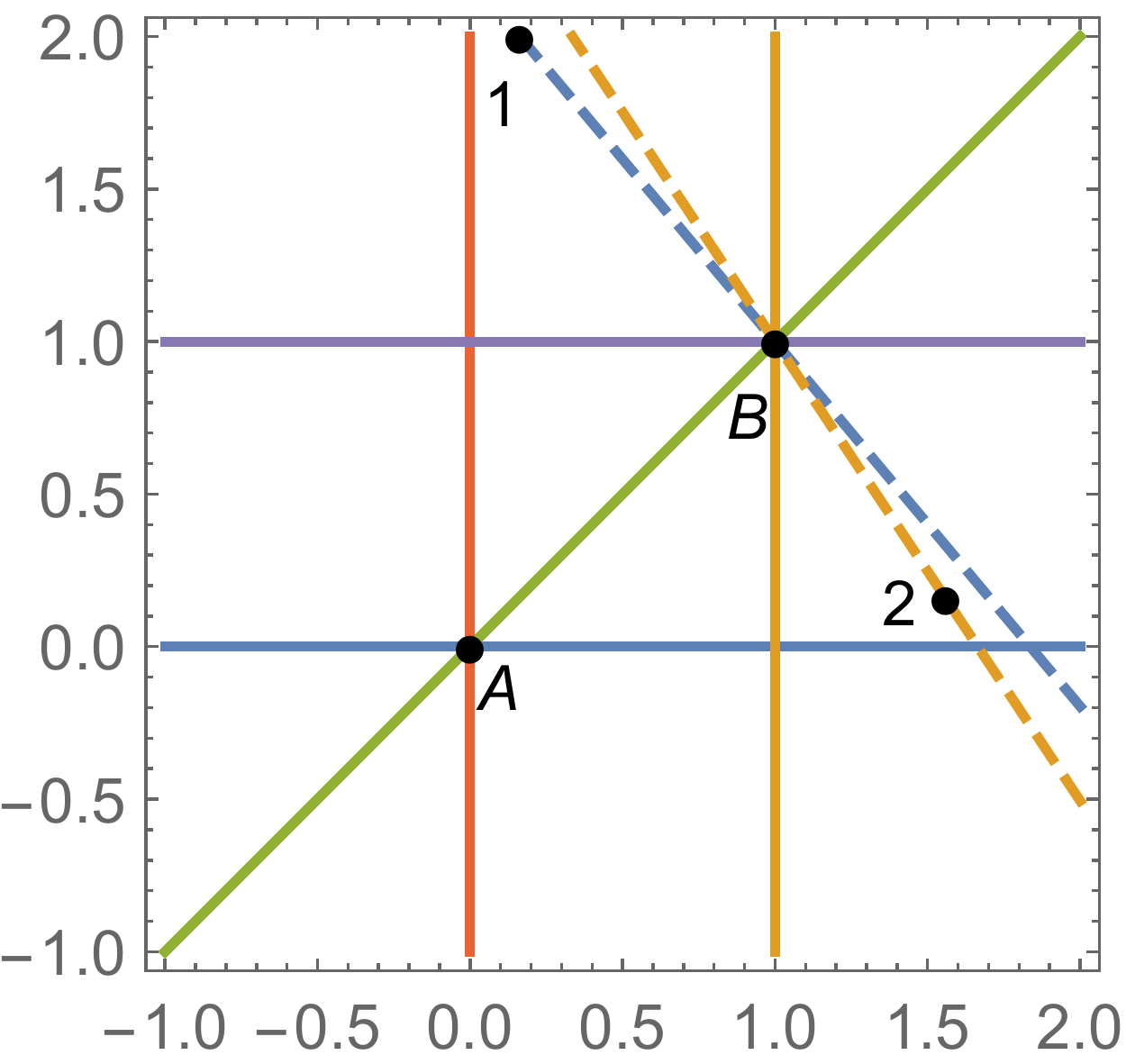}
\caption{\label{fig:generic}Three families of configurations of solutions on $\mathbb{RP}^2$ for generic positive kinematics. Left: Solutions in the same chamber. Center: Solutions in distinct chambers. Right: Solutions outside of the square $[0,1]^2$.}
\end{figure}

In fact, on kinematics that is positive but generic, i.e., not near any soft region, we find that the $26$ solutions split in classes shown in Figure~\ref{fig:generic}. In $16$ of them points $1$ and $2$ are located on the same $n=4$ bounded chamber while $6$ have $1$ and $2$ on two distinct bounded chambers. The remaining four have both $1$ and $2$ in unbounded chambers. This is possible because one particle generates a new bounded chamber for the other!
In the examples given in Figure~\ref{fig:generic} (right), point $2$ generates a bounded chamber for $1$ with the line $2{-}B$, while particle $1$ generates a bounded chamber for $2$ with the line $1{-}B$, both indicated by dashed lines.

Before ending this section let us mention that one can be tempted to continue the construction of chambers for more particles and count the number of solutions to the $k=3$ scattering equations in this way. More explicitly, placing particle $2$ in any of the $13$ bounded chambers generated by particle $1$ gives rise to a diagram with $42$ bounded chambers. This implies that there must be \emph{at least} $2\times 13\times 42$ solutions for $n=7$. In Appendix~\ref{app:soft-limits} we carefully study soft limits directly from the scattering equations and find a closed formula for the bounded regions and also explain why for $n>6$ and $k=3$ these numbers are strict lower bounds.

\section{\label{sec:discussion}Discussion}

In this work we have initiated the study of a natural generalization of scattering equations to moduli spaces of points on $\mathbb{CP}^{k-1}$. We have only scratched the surface of what seems to be a vast subject. The duality explained in Section \ref{sec:duality} between the space of $n$ points on $\mathbb{CP}^{k-1}$ and that of $n$ points on $\mathbb{CP}^{n-k-1}$ shows that the standard $k=2$ scattering equations, which are at the heart of the CHY formalism, already imply the existence of the $k=n{-}2$ scattering equations. Moreover, the standard Mandelstam invariants $s_{ab}$ give rise to $k=n{-}2$ kinematic invariants. In this work we have filled the gap between these two end points by studying intermediate values of $k$.

We found that the all the elements needed to define generalized CHY amplitudes are also present for any $k$ and gave the explicit formulation of ``biadjoint amplitudes''. The first non-trivial case is six points on $\mathbb{CP}^{2}$, i.e., the scattering equations relating the space of kinematic invariants $\s_{abc}$ to $X(3,6)$. Explicit computations of biajoint amplitudes led to the discovery of new kinematic poles $R$ and $\tilde R$ in addition to the expected $\s$ and $\t$ poles. A very pressing question is the explicit computation for $n>6$ and $k=3$. A direct computation for seven particles seems technically very challenging as the number of solutions jumps from $26$ for $n=6$ to more than a thousand for $n=7$. However, it is well known that in the $k=2$ case, a variety of techniques have been developed that allow the evaluation of CHY formulae without actually solving the scattering equations, see, e.g., \cite{Baadsgaard:2015voa,Gomez:2016bmv,Gao:2017dek}.

One of the early successes of the CHY formalism was the direct connection between biadjoint scalar amplitudes and the Kawai--Lewellen--Tye construction and the simple derivation of the Bern--Carrasco--Johansson basis of color ordered amplitudes. These developments rely on the fact that the basis of Parke--Taylor functions on $X(2,n)$ has size $(n{-}3)!$ when evaluated on the solutions of the scattering equations which happens to be the same as the number of solutions. It is clear that a generalization of these construction to higher $k$ is very desirable. At this point there is evidence that Parke--Taylor functions might not provide the most general basis of functions needed for $2<k<n-2$.

Another natural question is the generalization of the notion of Feynman diagrams for $k>2$. When $n=6$ and $k=3$ we have made a proposal based on the computations of Section~\ref{sec:scattering-equations} and on the surprising connection to tropical Grassmannians explored and used in Section~\ref{sec:tropical}. Tropical Grassmannians are polyhedral complexes with what seems to be a very direct connection to the space of kinematic invariants. The vertices of the tropical Grassmannian are in bijection with all possible poles in a $\phi^3$ theory. Moreover, the facets are in bijection with all possible individual Feynman diagrams of the theory. We have found that the same is true for $k=3$ and $n=6$ for the vertices and kinematic poles. Using the natural assumptions that facets must also correspond to Feynman diagrams we developed a combinatorial computation of all biadjoint amplitudes by computing the most fundamental one, $m_6^{(3)}(\mathbb{I}|\mathbb{I})$. We noted that one class, the EEEE class, of $k=3$ Feynman diagrams did not contribute to $m_6^{(3)}(\mathbb{I}|\mathbb{I})$ and hence to any of the biadjoint amplitudes. This is strong evidence that for $k=3$ one has to go beyond Parke--Taylor factors as building blocks for the integrand.

Another piece of evidence for the need of more general integrands comes from the $k=3$ Schouten identity. In $k=2$ the Schouten identity
\be\label{sch}
(13)(24)=(12)(34)+(14)(23)
\ee
gives rise what is known as the $U(1)$-decoupling identity among Parke--Taylor functions by dividing \eqref{sch} by all the factors
\be
\frac{1}{(12)(23)(34)(41)} + \frac{1}{(14)(42)(23)(31)}+\frac{1}{(12)(24)(43)(31)} = 0. 
\ee
The $k=3$ Schouten identity reads 
\be
(123)(456)+(341)(256) = (234)(156)+(412)(356).
\ee
In the same way as before this leads to the following $4$-term identity
\begin{align}\label{poss}
\nonumber \frac{1}{(341)(256)(234)(156)(412)(356)}+\frac{1}{(123)(456)(234)(156)(412)(356)} &= \\ \frac{1}{(412)(356)(123)(456)(341)(256)}+\frac{1}{(234)(156)(123)(456)(341)(256)}.
\end{align}
None of these four terms are Parke--Taylor functions. However, it is possible to multiply \eqref{poss} by a product of six factors, e.g., $(236)^2(145)^2(612)(345)$, such that each of the four terms are turned into the product of two Parke--Taylor functions but we take this still as an indication that more general functions are needed.

We also found an explicit realization of the $k$ space of kinematic invariants in terms of a generalization of the spinor-helicity formalism. Rank one $2\times 2$ matrices are replaced by rank one $k\times k$ matrices and invariants are computed as determinants of sums of such matrices. Since the factorization, $s_{ab}=\langle ab\rangle [ab]$, that led to miraculous simplifications for $k=2$ also happens for any $k>2$, it is tempting to suggest that there is a generalization of constructions such as BCFW recursion, superamplitudes, etc. One immediate challenge is the definition of the notion of helicity and polarization vectors. We leave these fascinating questions for future research.

Another interesting connection made for $k=2$ is that between the CHY formalism, string theory \cite{Gomez:2013wza,Siegel:2015axg,Casali:2016atr,Mizera:2017rqa,Mizera:2017sen,Casali:2017mss}, $Z$-theory \cite{Carrasco:2016ldy,Mafra:2016mcc,Carrasco:2016ygv}, and ambitwistor strings \cite{Mason:2013sva,Geyer:2016nsh}. It is clear that a generalization of $Z$-theory integrals is possible for $k>2$ and natural to expect that on a subregion of the positive kinematics defined in Section~\ref{sec:poskin}, where appropriate $R$'s are positive, such integrals should exist.

Finally, there are two more very useful constructions known for $k=2$ which can greatly impact the ability of solving $k>2$ scattering equations. The first is obtaining what is known as the polynomial form of the scattering equations first constructed by Dolan and Goddard in \cite{Dolan:2014ega}. These polynomial form allows a faster and more stable numerical search as well as a direct way of counting the number of solutions using Bezout's theorem since the equations increase their degree in steps of one. The second development is the identification of an integrable system with the scattering equations for a particular set of kinematics invariant. This was done by Kalousios and led to a connection to Jacobi polynomials \cite{Kalousios:2013eca}. We started the exploration of such kind of kinematics in Section~\ref{sec:poskin} but leave the search for an integrable system for future work. Either one of these construction would help in completing the counting of solutions started in Section~\ref{sec:poskin} and in Appendix~\ref{app:soft-limits} where an attempt to follow the soft limit approach led only to lower bounds.

\section*{Acknowledgements}

We would like to thank F. Borges, D. Garcia, J. Rojas, and S. Yusim for useful discussions. 
N.E. would like to thank Perimeter Institute for their support and hospitality while this work was initiated. A.G. thanks CONICYT for financial
support.
This research was supported in part by Perimeter Institute for Theoretical Physics. Research at Perimeter Institute is supported by the Government of Canada through the Department of Innovation, Science and Economic Development Canada and by the Province of Ontario through the Ministry of Research, Innovation and Science.

\renewcommand{\thefigure}{\thesection.\arabic{figure}}
\renewcommand{\thetable}{\thesection.\arabic{table}}
\appendix

\section{\label{app:Euler-characteristic}Euler Characteristic of $X(3,6)$}

In this appendix we give a derivation of the fact that scattering equations on $X(3,6)$ have $26$ solutions. As explained in \cite{Mizera:2017rqa}, CHY formulae can be understood as intersection numbers of twisted differential forms, which in our case belong to the cohomology group
\be\label{twisted-cohomology}
H^4(X(3,6),\, d{\cal S}_3 \wedge) := 
\frac{4\text{-forms on }X(3,6)}{d{\cal S}_3 \wedge (3\text{-forms on }X(3,6))},
\ee
where ${\cal S}_3$ is the Morse function \eqref{gen} for $n{=}6$. In other words, it is the space of $4$-forms modulo anything proportional to the scattering equations. This realization allows us compute the dimension of the space \eqref{twisted-cohomology}, and hence also the number of solutions ${\cal N}_6^{(3)}$ of the scattering equations \eqref{sck3}, purely topologically as the Euler characteristic of $X(3,6)$. More precisely, under the assumption that the kinematics is generic (so that all other twisted cohomology groups vanish), one can show that
\be\label{number-of-solutions}
{\cal N}_6^{(3)} = (-1)^4\, \chi(X(3,6))
\ee
using Morse theory, see, e.g., \cite{aomoto1975,Mizera:2017rqa}. We can use a chart in which punctures $\{1,2,3,4\}$ are held fixed, such that $X(3,6)$ can be written as the complement of the projective variety
\be
V := \bigcup_{\substack{1 \leq i < j < k \leq 6\\ k \geq 5}} \!\!\! \{ \,\la ijk \ra = 0\,\}
\ee
in $(\mathbb{CP}^2)^2$. Using the inclusion-exclusion principle we can write the Euler characteristic as
\begin{align}
\chi(X(3,6)) &= \chi((\mathbb{CP}^2)^2) - \chi(V)\nonumber\\
&=3{\times}3 - \chi(V),\label{Euler}
\end{align}
where we used $\chi(\mathbb{CP}^2) = 3$. The remaining contribution $\chi(V)$ can be evaluated using the algebraic geometry system \texttt{Macaulay2} \cite{M2} with the package \texttt{CharacteristicClasses.m2} \cite{jost2015computing,HELMER2017548} as follows. Let $\texttt{R}$ be the coordinate ring of $(\mathbb{CP}^2)^2$ (say over $\mathbb{Z}/p\mathbb{Z}$ for $p{=}32749$) with coordinates $\var{r_i}$ for $\mathtt{i}=0,1,\ldots,5$, and $\texttt{I}$ be the ideal generated by vanishing of all the relevant maximal minors of the matrix
\be
\left(
\begin{array}{cccccc}
	1 & 0 & 0 & 1 & \var{r_0} & \var{r_3} \\
	0 & 1 & 0 & 1 & \var{r_1} & \var{r_4} \\
	0 & 0 & 1 & 1 & \var{r_2} & \var{r_5} \\
\end{array}
\right).
\ee
Then the Euler characteristic in \eqref{Euler} can be computed using the following script:
\begin{verbatim}
load "CharacteristicClasses.m2";

R = MultiProjCoordRing(ZZ/32749, symbol r, {2,2});
I = ideal(r_0*r_1*r_2*r_3*r_4*r_5*(r_0-r_1)*(r_0-r_2)*(r_1-r_2)*(r_3-r_4)
          *(r_3-r_5)*(r_4-r_5)*(r_0*r_4-r_1*r_3)*(r_0*r_5-r_2*r_3)
          *(r_1*r_5-r_2*r_4)*(r_0*(r_4-r_5)+r_1*(r_5-r_3)+r_2*(r_3-r_4)));

3*3-Euler(I)
\end{verbatim}
The output of this computation is $26$, which by \eqref{number-of-solutions} gives the number of solutions of scattering equations, ${\cal N}_6^{(3)}=26$. Similar computation for $X(3,5)$ yields ${\cal N}_5^{(3)}=2$.

\section{\label{app:soft-limits}Soft Limits and Numbers of Solutions}

One of the most basic questions about the scattering equations is the number of solutions. It is well-known that the $k=2$ scattering equations for $n$ particles have $(n-3)!$ solutions \cite{Cachazo:2013gna}. Let us review one technique for proving it and then try and apply it for $k=3$. The idea is to approach what is known as the soft-limit region for the $n^{\rm th}$ particle. This is done by writing all invariants of the form $\s_{an}$ as $\tau \hat\s_{an}$ in the limit when $\tau\to 0$. In the limit the $n^{\rm th}$ particle drops from the first $n-1$ scattering equations which become those for a system of $n-1$ particles. The $n^{\rm th}$ equation is proportional to $\tau$ even for finite $\tau$ and therefore it fixes the location of $x_n$. Assuming that the system for $n-1$ has been solved and ${\cal N}_{n-1}^{(2)}$ solutions have been found, each such solution $x_{a,(I)}$ give rise to an equation for $x_n$
\be\label{sofy}
\sum_{a=1}^{n-1}\frac{\hat\s_{an}}{x_{a,(I)}-x_n} = 0.
\ee
It is easy to show that this leads to $n-3$ solutions for $x_n$. Therefore ${\cal N}_{n}^{(2)}=(n-3){\cal N}_{n-1}^{(2)}$. For $n=4$ \eqref{sofy} is all there is after fixing three points and therefore ${\cal N}_{4}^{(2)}=1$. This leads to the expected result ${\cal N}_{n}^{(2)} = (n-3)!$.

Strictly speaking, the soft argument only leads to a lower bound on the number of solutions as one has to prove that when taking $\tau=0$ in the first $n-1$ equations one is allowed to drop terms that depend on particle $n$. For instance, it could happen that there are solutions where $x_n- x_a\sim {\cal O}(\tau)$ and hence the term $s_{an}/(x_n-x_a)$ cannot be dropped. Indeed this happens when collinear limits, $s_{an}\to 0$, are taken and it is well-known that solutions split into two classes: singular and regular. Regular solutions are those for which the term $s_{an}/(x_n-x_a)$ can be dropped. This means that the soft limit argument is only guarantee to count the regular solutions. It turns out that when $k=2$ there are no singular solutions in the soft limit. Unfortunately, this is not the case when $k>3$. 

Here we repeat the same argument for $k=3$, one takes the ``soft" limit $\s_{abn}=\tau \hat\s_{abn}$ with $\tau\to 0$. We will remove all dependence on particle $n$ in the equations that defined the $(n-1)$ system, i.e., we will only compute the number of regular solutions. 

Assuming that the system for $(n-1)$ particles has been solved, one finds for each solution $x_{a,(I)},y_{a,(I)}$ two equations
\be\label{raq}
\sum_{1\leq b<c \leq n-1}\frac{\s_{nbc}(x_{b,(I)}-x_{c,(I)})}{\langle nbc\rangle_{(I)}}=0 ,\qquad  \sum_{1\leq b<c \leq n-1}\frac{\s_{nbc}(y_{b,(I)}-y_{c,(I)})}{\langle nbc\rangle_{(I)}}=0.
\ee
We are only interested in counting the number of solutions to these ``soft" equations. In order to analyze these equations one has to rewrite them as the ratio of two polynomials
\be\label{rati}
\frac{P_A(x_n,y_n)}{\prod_{1\leq b<c \leq n-1}\langle nbc\rangle_{(I)}} = 0, \qquad \frac{P_B(x_n,y_n)}{\prod_{1\leq b<c \leq n-1}\langle nbc\rangle_{(I)}} = 0.
\ee

Counting the number of solutions to the system $P_A(x_n,y_n)=P_B(x_n,y_n)=0$ is harder than in the $k=2$ case for two reasons.

In order to understand the first, recall that when dealing with a single polynomial in one variable, as in $k=2$, the degree of the polynomial equation coming from the numerator of \eqref{sofy} directly gives the number of solutions.
The first complication for $k=3$ arises from the fact that for two polynomials in two variables, $(x_n,y_n)$, Bezout's theorem only gives an upper bound for the number of solutions at finite locations. The bound is the product of the degrees of the two polynomials.

The second difficulty comes from the fact that the system $P_A(x_n,y_n)=P_B(x_n,y_n)=0$ has solutions which are not solutions to the original equations \eqref{raq}. Such solutions come from where two factors in the denominator of \eqref{rati} vanish. To see this note that if the poles are approached as $\epsilon\to 0$, then the original equations \eqref{raq} diverge as $1/\epsilon$ while the denominators in \eqref{rati} diverge as $1/\epsilon^2$. This shows that both $P_A$ and $P_B$ must vanish as $\epsilon\to 0$. This means that we have to remove these spurious solutions from Bezout's bound.

Very nicely, the final formula turns out to be simple
\be\label{n3}
{\rm Soft}^{(3)}_n = \frac{1}{8}(n-4)(n^3-6n^2+11n-14).
\ee
Now it is tempting to construct the number of solutions as the product of all soft factors up to the number of particles of interest, i.e.
\be
{\cal N}_{n}^{(3):{\rm naive}} =\prod_{m=5}^n {\rm Soft}^{(3)}_m.
\ee
However, as discussed above, the soft limit computation only captures regular solutions. We will now show why when $k=3$ there must also be singular solutions as well and leave their enumeration for future work. 

Note that ${\cal N}_{n}^{(3):{\rm naive}}$ seems to contain a factor of $(n-4)!$. However, since the polynomial $(m^3-6m^2+11m-14)$ is not divisible by $8$ for all $m$ one cannot conclude that ${\cal N}_{n}^{(3):{\rm naive}}$ is divisible by $(n-4)!$. In fact already for $n=8$, this naive number of solutions is not divisible by $(n-4)!$ which is a contradiction with the $(3,n)\Leftrightarrow (n-3,n)$ duality discussed in Section~\ref{sec:duality}.

To see this more explicitly, note that when $n=7$ one can repeat the same soft argument for $k=4$. Starting from $n=6$, $k=2$ which has six solutions one finds that so does $n=6$, $k=4$ by the duality in Section~\ref{sec:duality}. An explicit computation reveals that the soft equations for the seventh particle have $192$ solutions. This means that the number of solutions must be at least $6 \times  192= 1152$. This is $60$ more than ${\cal N}_{7}^{(3):{\rm naive}}=2\times 13\times 42= 1092$.

\section{\label{app:numerics}Numerical Solution to the $X(3,6)$ Scattering Equations}

The $k=3, n=6$ scattering equations have $26$ solutions. For general kinematic invariants the 26 solutions cannot be found analytically. In this appendix we give and explicit procedure for finding all $26$ solutions to arbitrarily-high precision for a kinematic point in the positive region. The procedure involves finding approximate solutions with low precision, which we call {\it seeds}, using a numerical search and then each seed is used as an input in the Mathematica's \texttt{FindRoot} function set up to high precision.

One the kinematic points we used is given by the following values of the independent invariants
\begin{gather}
\s_{123} = 139,\quad \s_{124} = 179,\quad \s_{125} = 223,\quad 
\s_{126} = 257,\quad \s_{134} = 199,\nn\\
\s_{135} = 241,\quad 
\s_{136} = 281,\quad \s_{145} = 271,\quad \s_{146} = 313,\quad 
\s_{156} = -2103,\nn\\
\s_{234} = 227,\quad \s_{235} = 263,\quad 
\s_{236} = 307,\quad \s_{245} = 283,\quad \s_{246} = 337,\\ 
\s_{256} = -2215,\quad \s_{345} = -2025,\quad \s_{346} = -2239,\quad 
\s_{356} = 2607,\quad \s_{456} = 2455.\nn
\end{gather}
It is easy to verify that these invariants satisfy momentum conservation and that none of the possible poles of $k=3$ Feynman diagrams vanish.

We use the same gauge fixing as in Section~\ref{sec:scattering-equations}:
\be
\left(
\begin{array}{cccccc}
	1 & 1 & 1 & 1 & 1 & 1 \\
	0 & 1 & 0 & 1 & x_5 & x_6 \\
	0 & 0 & 1 & 1 & y_5 & y_6 \\
\end{array}
\right).
\ee
The seeds for the $26$ solutions are
\be
\arraycolsep=0.9em
\begin{array}{llll}
 x_5\to -1.81336 & x_6\to 2.63087 & y_5\to
   2.72136 & y_6\to -1.06766 \\
 x_5\to 2.44712 & x_6\to -1.79167 & y_5\to
   -0.903713 & y_6\to 2.69014 \\
 x_5\to 0.534746 & x_6\to 0.537102 & y_5\to
   -2.70181 & y_6\to -0.201572 \\
 x_5\to 0.534421 & x_6\to 0.528164 & y_5\to
   -0.175607 & y_6\to -2.36545 \\
 x_5\to 1.8782 & x_6\to -3.70997 & y_5\to
   1.82661 & y_6\to -2.60442 \\
 x_5\to 0.50317 & x_6\to 0.470842 & y_5\to
   0.374444 & y_6\to -1.38027 \\
 x_5\to 1.04874 & x_6\to 1.55304 & y_5\to
   0.24654 & y_6\to -0.45775 \\
 x_5\to 0.475295 & x_6\to 0.503754 & y_5\to
   -1.54168 & y_6\to 0.365724 \\
 x_5\to 0.237534 & x_6\to 0.836314 & y_5\to
   1.70422 & y_6\to 1.57173 \\
 x_5\to -2.92233 & x_6\to 1.88376 & y_5\to
   -1.97545 & y_6\to 1.82657 \\
 x_5\to 1.01926 & x_6\to 1.16625 & y_5\to
   0.661554 & y_6\to 0.389534 \\
 x_5\to 1.00838 & x_6\to 0.917236 & y_5\to
   1.08258 & y_6\to 0.909572 \\
 x_5\to 1.51702 & x_6\to 1.05038 & y_5\to
   -0.425664 & y_6\to 0.257 \\
 x_5\to 0.386006 & x_6\to 0.653472 & y_5\to
   0.712486 & y_6\to 0.731654 \\
 x_5\to 0.500766 & x_6\to 0.49462 & y_5\to
   0.420314 & y_6\to 0.142516 \\
 x_5\to -0.00949757 & x_6\to 0.094872 & y_5\to
   1.11668 & y_6\to 0.896041 \\
 x_5\to 0.9293 & x_6\to 1.00989 & y_5\to 0.92216
   & y_6\to 1.10112 \\
 x_5\to -0.697705 & x_6\to -0.0509408 & y_5\to
   -0.583952 & y_6\to 0.266517 \\
 x_5\to 0.841712 & x_6\to 0.242035 & y_5\to
   1.58999 & y_6\to 1.68593 \\
 x_5\to 0.0782592 & x_6\to -0.0105854 & y_5\to
   0.913617 & y_6\to 1.14256 \\
 x_5\to -0.0198028 & x_6\to -0.188344 & y_5\to
   0.63874 & y_6\to 0.367487 \\
 x_5\to 1.17891 & x_6\to 1.02194 & y_5\to
   0.372119 & y_6\to 0.662613 \\
 x_5\to -0.203209 & x_6\to -0.0221714 & y_5\to
   0.351966 & y_6\to 0.642902 \\
 x_5\to -0.0501271 & x_6\to -0.755866 & y_5\to
   0.254807 & y_6\to -0.636579 \\
 x_5\to 0.650306 & x_6\to 0.383491 & y_5\to
   0.727378 & y_6\to 0.716881 \\
 x_5\to 0.495514 & x_6\to 0.500968 & y_5\to
   0.128367 & y_6\to 0.41208 \\
\end{array}
\ee
Using the function \texttt{FindRoot} with four scattering equations as arguments and the seeds as initial conditions one can easily generate all $26$ solutions to $500$ digits of precision. 

Using the high-precision solutions the $k=3$ biadjoint formulas can be evaluated. The numerical result, which is guaranteed to be a rational number since the coefficients of all scattering equations are rational, can be easily turn into a rational number using the Mathematica function \texttt{Rationalize}.

\bibliographystyle{JHEP}
\bibliography{references}

\end{document}